\documentclass[11pt,a4paper,x11names]{article}
\usepackage{jheppub}

\usepackage{amsfonts}
\usepackage{tikz-cd,tikz}
\usepackage{comment}
\usepackage{physics}
\usepackage{float}
\usepackage{listings}
\usepackage{calc}

\usepackage{tikz}
\usepackage{subcaption}

\DeclareFixedFont{\ttb}{T1}{txtt}{bx}{n}{12}
\DeclareFixedFont{\ttm}{T1}{txtt}{m}{n}{12}
\DeclareFixedFont{\ttms}{T1}{txtt}{m}{n}{9}
\DeclareFixedFont{\ttmss}{T1}{txtt}{m}{n}{7}

\definecolor{deepblue}{rgb}{0.0, 0.0, 0.55} 
\definecolor{shadecolor}{rgb}{0.96,0.96,0.91}
\definecolor{shadecolor2}{rgb}{1,1,0.99}

\newcommand{\mcomment}[1]{}
\newcommand\mathstyle{\lstset{
language=Mathematica,
basicstyle={\scriptsize\def\fvm@Scale{.5}\fontfamily{fvm}\selectfont},
otherkeywords={self},
keywordstyle=\ttb\scriptsize\color{deepblue},
emph={MyClass,__init__},
emphstyle=\ttb\color{deepred},
backgroundcolor=\color{pink!20!white},
stringstyle=\color{deepgreen},
commentstyle=\color{SkyBlue3!70!PaleGreen4},
frame=tb,
showstringspaces=false
}}

\newwrite\todofile
\immediate\openout\todofile=\jobname.tdo

\newcounter{todocounter}

\newcommand{\printtodos}{
        \section*{To-Do List}
        \immediate\closeout\todofile
        \input{\jobname.tdo}
}
\lstnewenvironment{mathematica}[1][]
{
\mathstyle
\lstset{#1}
}
{}

\newcommand{\la}[1]{\label{#1}}
\newcommand{\eq}[1]{(\ref{#1})}
\newcommand{\nn}{\nonumber}

    \newcommand{\beq}{\begin{equation}}
    \newcommand{\eeq}{\end{equation}}
    \newcommand\beqa{\begin{eqnarray}}
    \newcommand\eeqa{\end{eqnarray}}
\newcommand\bea{\begin{array}}
\newcommand\eea{\end{array}}

\newcommand{\bm}{{\bar m}}
\newcommand{\bn}{{\bar n}}

\newcommand{\ii}{i}

\newcommand{\algsu}{\mathfrak{su}}

\newlength{\widthLOne}
\newlength{\widthLTwo}
\newcommand{\bfT}{\mathbf{T}}

\newcommand{\sx}{\mathsf{x}}
\newcommand{\sy}{\mathsf{y}}
\newcommand{\bsx}{\bar{\mathsf{x}}}
\newcommand{\Wr}{\text{Wr}}
\newcommand{\btheta}{\overline{\theta}}

\newcommand{\bQ}{\mathbf{Q}}

\newcommand{\sfP}{\mathsf{P}}

\newcommand{\svy}{\mathsf{y}}
\newcommand{\svx}{\mathsf{x}}
\newcommand{\lM}{\mathcal{M}}
\newcommand{\sP}{\textsf{P}}
\newcommand{\lO}{\mathcal{O}}

\newcommand{\bl}{\Big(\hspace{-1.85mm}\Big(\,}
\newcommand{\br}{\,\Big)\hspace{-1.85mm}\Big)}

\newcommand{\bbE}{\mathbb{E}}

\newcommand{\gl}{\mathfrak{gl}}
\newcommand{\su}{\mathfrak{su}}

\newcommand{\sfe}{\mathsf{e}}
\newcommand{\lE}{\mathcal{E}}
\newcommand{\lL}{\mathcal{L}}
\newcommand{\lN}{\mathcal{N}}

\newcommand{\T}{\mathbb{T}}

\def\[{\left[}
\def\]{\right]}
\def\({\left(}
\def\){\right)}
\def\<{\left<}
\def\>{\right>}

\title{Boundary Overlaps from Functional Separation of Variables}
\author{Simon Ekhammar, Nikolay Gromov and Paul Ryan}
\affiliation{
Department of Mathematics, King's College London
}\emailAdd{simon.ekhammar@kcl.ac.uk}
 \emailAdd{nikolay.gromov@kcl.ac.uk}
 \emailAdd{paul.1.ryan@kcl.ac.uk}

\abstract{
In this paper we show how the Functional Separation of Variables (FSoV) method can be applied to the problem of computing overlaps with integrable boundary states in integrable systems.
We demonstrate our general method on the example of a particular boundary state, a singlet of the symmetry group, in an $\mathfrak{su}(3)$ rational spin chain in an alternating fundamental--anti-fundamental representation. The FSoV formalism allows us to compute in determinant form not only the overlaps of the boundary state with the eigenstates of the transfer matrix, but in fact with any factorisable state. This includes off-shell Bethe states, whose overlaps with the boundary state have been out of reach with other methods. Furthermore, we also found determinant representations for insertions of so-called Principal Operators (forming a complete algebra of all observables) between the boundary and the factorisable state as well as certain types of multiple insertions of Principal Operators. Concise formulas for the matrix elements of the boundary state in the SoV basis and $\mathfrak{su}(N)$ generalisations are presented. Finally, we managed to construct a complete basis of integrable boundary states by repeated action of conserved charges on the singlet state. As a result, we are also able to compute the overlaps of all of these states with integral of motion eigenstates.
}

\date{September 2023}

\begin{document}

\maketitle

\newpage

\section{Introduction}

Functional Separation of Variables (FSoV) is an integrability tool initially developed in ${\cal N}=4$ SYM and used to compute expectation values of cusped Wilson loops directly in the language of Q-functions, the main objects of the Quantum Spectral Curve (QSC) formalism~\cite{Cavaglia:2018lxi}. Since then, it has been developed for integrable spin chains of arbitrarily high rank $\mathfrak{sl}(N)$~\cite{Cavaglia:2019pow,Gromov:2019wmz,Gromov:2020fwh}, conformal fishnet theory \cite{Cavaglia:2021mft}, and correlators of single-trace local operators in planar $\lN=4$ SYM at weak coupling~\cite{Bercini:2022jxo}, and seems to be merging well with the form-factor approach \cite{Basso:2022nny}. In spin chains, the state-of-the-art is the computation from FSoV of correlation functions of a complete set of observables, known as Principal Operators~\cite{Gromov:2022waj}.

In this paper we will apply the FSoV method to the computation of a different class of observables in integrable systems – overlaps $\langle B|\Psi\rangle$, where $|\Psi\rangle$ is a Hamiltonian eigenstate, or more generally a factorisable state in the separation of variables (SoV) framework, and $\langle B|$ is a so-called integrable boundary state -- a state annihilated by half of the conserved charges of the model. Such overlaps have attracted a huge amount of attention in recent years across the high-energy theory and condensed matter communities. In particular, in the context of AdS/CFT due to their relation to one-point functions in defect $\lN=4$ SYM \cite{deLeeuw:2015hxa,BuhlMortensen:2015gfd,deLeeuw:2017cop,DeLeeuw:2018cal,deLeeuw:2019usb,Linardopoulos:2020jck,Komatsu:2020sup,Kristjansen:2023ysz} and ABJM theory \cite{Kristjansen:2021abc,Gombor:2022aqj}, three-point functions involving determinant operators and more generally g-functions or boundary entropy in 2D QFTs~\cite{Jiang:2019zig,Jiang:2019xdz,Caetano:2020dyp,Caetano:2021dbh,Yang:2021hrl}. Additionally, they have also been used to model quantum quenches~\cite{Fioretto:2009yq,pozsgay2014correlations,Piroli:2017sei,Piroli:2018ksf,piroli2019integrable} and are closely related to the representation theory of quantum groups with open boundary conditions \cite{Gombor:2021uxz,pozsgay2019integrable}.

Due to their prevalence, there has been substantial interest in computing these overlaps \cite{deLeeuw:2016umh,DeLeeuw:2019ohp}, for example through the nested Bethe ansatz approach, see in particular \cite{Gombor:2021hmj,Gombor:2022deb,Gombor:2023bez}.  In these approaches, the result for the overlap is expressed as a determinant built from Bethe roots – the so-called Gaudin determinant. Whereas it is very compact and easy to evaluate, the expressions in terms of Bethe roots have several limitations. First, for higher rank, these expressions are only valid for on-shell states.
Second, Bethe roots do not provide a complete characterisation of states at finite-coupling in planar $\lN=4$ SYM. Instead, states are characterised by Q-functions solving the Quantum Spectral Curve equations. Since Q-functions resolve both of these difficulties, we are motivated to consider overlaps $\langle B|\Psi\rangle$ directly in this language\footnote{Such results have already \cite{Gombor:2021uxz} been obtain in $\mathfrak{su}(2)$-based models using Sklyanin's formulation of SoV \cite{Sklyanin:1984sb}. The generalisation to $\mathfrak{su}(N)$ is highly non-trivial, but the FSoV technique considered here provides a technical advantage giving a shortcut to the result.}.
We  now summarise our main results. 

\paragraph{Distinguished boundary state.} First, we find that there is a unique distinguished integrable boundary state $\bra{W}$ which is naturally encoded in FSoV. This state is singled out by the property that it is a singlet of the global symmetry group. For this state, the overlap with an eigenstate $\ket{\Psi}$ of the transfers matrices (commuting conserved quantites) takes the following very concise determinant form in the case of $\mathfrak{su}(3)$ spin chains:
\begin{equation}\label{eqn:mainform}
    \langle W|\Psi\rangle = \det_{(a,\alpha),(s,\beta)} 
    \oint \frac{du}{2\pi \ii} {
     {\left(u+\frac{i\,s}4\right)^{2\beta-1}}\bQ_1(u)}\bQ_a(-u-\tfrac{i\,s}{2}) \rho_{\alpha}(u)\,.
\end{equation}
The rows and columns of the matrix are labelled lexicographically by the pairs $(a,\alpha)$ and $(s,\beta)$ respectively, with $a=1,3\,,s=1,-1\,,\alpha,\beta=1,\dots,L$ with $2L$ being the length of the spin chain. The contour of integration in \eqref{eqn:mainform} is a large circle enclosing all (finitely-many) poles of the integrand. $\bQ_a(u)$ are conveniently normalized Baxter Q-functions, whose zeroes are Bethe roots as defined in the main text and with poles at $\theta_{\alpha}\pm \frac{i}{2}$ where $\theta_\alpha$ are the spin chain inhomogeneities. The $i$-periodic functions $\rho_\alpha(u)$ are defined by
\beq\la{rhoalpha}
\rho_\alpha(u)=\frac{1}{\cal R}\prod_{\beta\neq \alpha}^L (1+e^{2\pi (u-\theta_\beta)})
\prod_{\beta=1}^L (1-e^{-2\pi (u+\theta_\beta)})\;,
\eeq
and the constant ${\cal R}$ is such that $\rho_\alpha(\theta_\alpha+\tfrac{i}{2})=1$.

\paragraph{Insertions.}

We also find similar determinant expressions for a large class of related observables. These are of the form $\langle W|\mathsf{P}_{\pm,r}^\gamma(v)|\Psi\rangle$, with indices $r\in \{\pm 3,\pm 1\}$ with $\gamma$ and $v$ being generic complex numbers. $\mathsf{P}^\gamma_{\pm,r}(v)$ are essentially the Principal Operators introduced in \cite{Gromov:2022waj} -- a special class of operators acting on the spin chain Hilbert space and generating the full algebra (Yangian) of observables -- albeit with slight modifications for the boundary set-up. Note that the states $\langle W|\mathsf{P}^\gamma_{\pm,r}(v)$ are in general not integrable boundary states, but nevertheless their overlaps are naturally computable by FSoV. Additionally, we find similar determinant representations for certain multiple insertions of these operators. 

\paragraph{Overlap with a complete basis of integrable boundary states.}

We found that the state $\langle W|$ has another remarkable feature -- by repeated action of integrals of motion on it, we found that we can generate a basis in the space of integrable boundary states. More concretely, we found that \textit{any} integrable boundary state $\langle B|$ of our system admits a linear expansion into basis integrable states $\langle B_n|$, where $n$ is a tuple $n=(n_1,\dots,n_L)$ and $n_\alpha=0,1,2$, defined as 
\begin{equation}
    \langle B_n|\, \propto \, \langle W|\displaystyle \prod_{\alpha=1}^L \left(\hat{\tau}_+(\theta_\alpha+i/2)\right)^{n_\alpha}
\end{equation}
where $\hat{\tau}_+$ is the fundamental transfer matrix of the system. This result is highly reminiscent of the construction of Separation of Variables bases \cite{Maillet:2018bim,Ryan:2018fyo,Ryan:2020rfk}. This construction with transfer matrices allows us to easily compute all overlaps $\langle B_n|\Psi\rangle$ as 
\begin{equation}\label{basisbdyintro}
   \langle B_n|\Psi\rangle =\langle W|\Psi\rangle\displaystyle \prod_{\alpha=1}^L \left(\frac{Q_1(\theta_\alpha-i/2)}{Q_1(\theta_\alpha+i/2)}\right)^{n_\alpha}\,,
\end{equation}
where $Q_1$ is the (twisted) polynomial part of $\bQ_1$, to be defined precisely in the main text \eqref{notation}. Schematically, this means that any overlap $\langle B|\Psi\rangle$ admits the following schematic form 
\begin{equation}
    \langle B|\Psi\rangle = \langle W|\Psi\rangle \times \text{boundary-dependent}
\end{equation}
where ``boundary-dependent" denotes a linear combination of terms of the form \eqref{basisbdyintro}, with coefficients fixed by the specific boundary.  Structurally, this result is very similar to overlaps which have been computed using Bethe ansatz approaches \cite{Gombor:2021hmj,Gombor:2023bez}, where the overlap takes the form of a ``boundary-dependent" piece times a ``universal" part, in that case related to the Gaudin determinant. This structure is also similar to what is found in integrable field theories \cite{Caetano:2020dyp}. It is noted in those papers however that the result is derived for on-shell Bethe states. In our formulation, the result also holds for off-shell states or any ``factorisable" state in the separation of variables approach. 

\paragraph{Paper outline.} 
The paper is organized as follows. In Section \ref{sec:def_system}, we introduce the definitions and main notations for the $\mathfrak{su}(3)$ spin chain we consider and introduce its Baxter equation. Section \ref{sec:crash_course_fsov} provides a comprehensive introduction to the Functional Separation of Variables (FSoV) method, including computation of the scalar product in determinant form, its SoV interpretation and the derivation of the SoV measure.
In Section \ref{sec:WPsiSection}, we explore the role of $\mathbb{Z}_2$ reflection symmetry, define the singlet boundary state and apply the FSoV technique to computing its overlaps with the eigenstates of the transfer matrices, or more generally any factorisable state, that is any state whose wave function factorises in the SoV bases, defined in the main text. Section \ref{sec:fsov_insertions} discusses the computation of overlaps with operator insertions within the FSoV framework, extending the analysis carried out in \cite{Gromov:2022waj} to integrable boundary states. 
Section \ref{sec:gen_integrable_states} is dedicated to generalising the overlaps to a complete basis of all integrable boundary states. We conclude with an Outlook, where we suggest potential future research directions. 

Numerous appendices provide in-depth supplementary material. Appendix \ref{app:MicroscopicDef} presents the microscopic formulation for $\su(3)$, detailing the Lax operators and RTT-formalism. Appendix \ref{app:sov} discusses the SoV bases, elaborating on the eigenstates of the operators $\mathbf{B}$ and $\mathbf{C}$, and explains how the results derived in the paper extend to those involving arbitrary factorisable states. 
In Appendix \ref{app:properties_w}, we explore the properties of the state $\langle W|$ in detail, and how it fits into the framework of integrable two-site states and the KT-relation. 
Appendix \ref{app:sigmification} focuses on the SoV representation of the various determinants arising from FSoV. The generalization of our results to $\mathfrak{su}(N)$ is treated in 
Appendix \ref{app:su_n_generalization}, while 
Appendix \ref{app:basis_integrable_states} details the construction of a basis for integrable boundary states. Finally, 
Appendix \ref{app:proofs_multiple_insertions} provides additional technical details regarding insertions.

\section{Q-functions for an Alternating $\su(3)$ Spin Chain}
\la{sec:def_system}

Instead of following the usual approach to integrable spin chains, beginning with a ``microscopic" realisation of the system in terms of Lax operators, we will jump straight to writing down the Q-system, Bethe Ansatz equations and Baxter TQ relations for our system. The traditional description in terms of Lax operators is relegated to Appendix \ref{app:MicroscopicDef}. 

\begin{figure}
    \begin{center}
\begin{tikzpicture}[scale=1.3,transform shape]
    \node[] (n1) at (-1,0) {};
    \node[] (n2) at (0,0) {};
    \node[] (n3) at (1,0){};
    \node[] (n4) at (2,0){};
    \node[] (ai1) at (-1,0){};
    \node[] (ao1) at (2,0){};
    \node[] (pi1) at (-1,-1){};
    \node[] (pi2) at (0,-1){};
    \node[] (pi3) at (1,-1){};
    \node[] (po1) at (-1,1){};
    \node[] (po2) at (0,1){};
    \node[] (po3) at (1,1){};
    \draw[thick,blue] (1.75,0) arc (-60:240:3cm and 0.2cm);
    \draw[thick,blue] (ai1)--(ao1);
    \filldraw[black,thick,fill=white] (n1) circle (0.28);
    \draw[black,thick,fill=white] (n2) circle (0.28);
    \draw[black,thick,fill=white] (n3) circle (0.28);
    \draw[black,thick,fill=white] (n4) circle (0.28);
    \node[color=magenta,thick,yshift=0,xshift=0.0] (a1) at (n1) {$\Box$};
    \node[color=magenta,thick,yshift=0.6] (a2) at (n2) {$\overline{\Box}$};
    \node[color=magenta,thick,yshift=0,xshift=0] (a3) at (n3) {$\Box$};
    \node[color=magenta,thick,yshift=0.6] (a4) at (n4) {$\overline{\Box}$};
    \node[anchor=north,yshift=-7] (t1) at (n1) {$\textcolor{red}{\theta_1}$};
    \node[anchor=north,yshift=-7] (t2) at (n2) {$\textcolor{red}{-\theta_1}$};
    \node[anchor=north,yshift=-7] (t3) at (n3) {$\textcolor{red}{\theta_2}$};
    \node[anchor=north,yshift=-7] (t4) at (n4) {$\textcolor{red}{-\theta_2}$};
\end{tikzpicture}    
\end{center}
    \caption{An alternating spin chain for $L=2$.}
    \label{fig:L2Cartoon}
\end{figure}
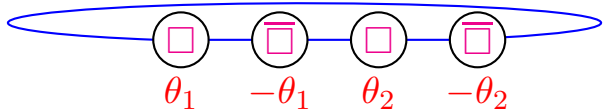

Let us explain the concrete model we aim to study. We consider a rational $\mathfrak{su}(3)$ spin chain of length $2L$, whose sites alternate between the fundamental and anti-fundamental representations of $\mathfrak{su}(3)$, with quasi-periodic boundary conditions. We will attach to each fundamental site an inhomogeneity $\theta_{\alpha}$ and to each anti-fundamental site $-\theta_{\alpha}$. A picture of the system for $L=2$ is given in Figure~\ref{fig:L2Cartoon}. This particular set-up is chosen to ensure that our system has enough symmetry to allow for integrable boundary states.

Throughout this paper we will use various types of indices. These are summarised in Table \ref{indextable}.

\begin{table}
  \centering
  \begin{tabular}{c|c}
    \hline
    \textbf{Index type} & \textbf{Range} \\
    \hline
    $\alpha,\beta$ & $1,\dots, L$ \\
    \hline
    $s,t$ & $\pm$ \\
    \hline
        $a,b$ & $1,2,3$ \\
    \hline
        $p$ & $1,\dots,2L$ \\
    \hline
        $n$ & $0,\dots,2L-1$ \\
    \hline
  \end{tabular}\caption{Common indices appearing in the paper and their range.}\label{indextable}
\end{table}

\subsection{Q-functions and Bethe Equations}\label{sec:QFunctionsSU3}

\paragraph{Q-system.}

Q-functions provide the most effective way to encode the conserved charges. Since we are working with an $\algsu({3})$ spin chain there are $3$ basic Q-functions, $Q_a(u)$, $a=1,2,3$, as well as their duals $Q^a(u)$, with $u$ a complex number known as the spectral parameter. 

For the model at hand \cite{Frassek:2011aa}, the Q-functions are \textit{twisted polynomials} parameterised as
\begin{equation}\la{Qnorm}
    Q_{a} = \lambda_{a}^{\ii u} \, q_{a}(u)\;, \quad 
    Q^a = \lambda_a^{-\ii u} \, q^a(u)\;,
\end{equation}
where $q_{a}$ and $q^a$ are monic polynomials in $u$ and $\lambda_1,\lambda_2,\lambda_3$ are twist parameters, encoding the quasi-periodic boundary conditions of the model.

The Q-functions obey QQ-relations. To write them down, it will be convenient to introduce some notation. First, we introduce standard notations for shifts of the spectral parameter 
\begin{equation}
    f^\pm :=f\left(u\pm \frac{i}{2} \right),\quad f^{[n]} :=f\left(u+ \frac{i\,n}{2} \right)\,.
\end{equation}
With this, we define the ``quantum Wronskian" $\Wr(f,g)$ of two functions $f$ and $g$ by
\begin{equation}
    \Wr(f,g):= f^+ g^- - f^- g^+\,.
\end{equation}
Next, we introduce ``source terms" $Q_{\theta},Q_{\btheta}$ \footnote{The ``bar" notation, e.g. $\bar\theta$, used here and elsewhere in the text never denotes complex conjugation. Instead, bar denotes ``reflected" objects, made precise in the language of Lax operators in Appendix \ref{app:MicroscopicDef}.}

\begin{equation}
    Q_{\theta}(u)= \prod_{\alpha=1}^{L}(u-\theta_{\alpha})\;,\quad
    Q_{\btheta}(u) = \prod_{\alpha=1}^{L}(u-\btheta_{\alpha})\;,\quad \bar{\theta}_\alpha:=-\theta_\alpha\;,
\end{equation}
encoding the inhomogeneities $\theta_\alpha\in \mathbb{C}$ of the spin chain. The split into two source terms is due to the fact that we are considering an alternating spin chain.

The QQ-relations then take the concise form
\begin{equation}\label{eqn:QQrelns}
    \Wr(Q_{1},Q_{2}) \ \propto (\lambda_1\lambda_2\lambda_3)^{i u}\, Q_{\theta}\,Q^3 \; ,
    \quad
    \Wr(Q^{1},Q^{2}) \ \propto (\lambda_1\lambda_2\lambda_3)^{-i u}\, \ Q_{\btheta}\, Q_3\;,
\end{equation}
with further relations obtained by cyclically permuting the indices $1,2,3$ and $\propto$ denotes equality up to $u$-independent factors, which are fixed by our normalization convention \eq{Qnorm}.

To classify all eigenstates of the system of conserved charges we need to solve the Q-system for all admissible choices of Q-functions. The possibilities can be classified using asymptotics, at large $u$ we have
\begin{equation}
    Q_a \sim \lambda_a^{i u} u^{M_a},\quad Q^a \sim \lambda_a^{-i u}u^{M^a}\,,
\end{equation}
where $M_a$ and $M^a$ are non-negative integers. By solving the QQ-relations at large $u$, we see in particular that 
\begin{equation}\la{wr2}
    M_1 + M_2 = L + M^3
\end{equation}
as well as cyclic permutations of $1,2,3$. Without loss of generality, we can take $M_1,M^3\leq L$. Once $M_{1}$ and $M^3$ have been specified the degrees of all other Q-functions are uniquely fixed. Furthermore, let us mention the $3\times 3$ Wronskian relation
\beq
\displaystyle\det_{a=1,2,3,\; z=-1,0,1} Q_a(u+ i z) \propto (\lambda_1\lambda_2\lambda_3)^{\ii u}\,Q_\theta^{+} Q_{\theta}^- Q_{\btheta}\;,
\eeq
which also implies
\beq\la{wr3}
M_1+M_2+M_3=3L\;.
\eeq
\paragraph{Bethe Ansatz equations.}

The Q-system implies the Bethe equations of the system in a standard way \cite{Kazakov:2015efa}. Denoting by $v_1$ a root of $Q_1$ and $v^3$ a root of $Q^3$ and by evaluating the QQ-relations at these roots we obtain
\begin{align}
    & \frac{Q_\theta(v_1-i/2)}{Q_\theta(v_i+i/2)} \frac{Q_1(v_1+i)}{Q_1(v_1-i)}\frac{Q^3(v_1-i/2)}{Q^3(v_1+i/2)} = -\frac{1}{\lambda_1\lambda_2\lambda_{3}}\;,\\
    & \frac{Q_{\bar\theta}(v^3-i/2)}{Q_{\bar\theta}(v^3+i/2)} \frac{Q^3(v^3+i)}{Q^3(v^3-i)}\frac{Q_1(v^3-i/2)}{Q_1(v^3+i/2)} = -{\lambda_1\lambda_2\lambda_{3}} \,.
\end{align}

Note that we have chosen to write the Bethe Ansatz equations in terms of $Q_1$ and $Q^3$. This is because we have implicitly picked a highest-weight state of our system defined by $q_1=q^{3}=1$. Of course, we are free to choose any highest-weight state, but this would necessitate a change of Q-functions. When we later discuss integrable boundary states we have also found this choice of Q-functions to be more natural in that context. 

\subsection{Baxter Equations}

The Baxter equation and its dual, analogues of the one-dimensional Schr\"odinger equation for the spin chain, are third order finite difference equations whose solutions are the various Q-functions. They can be compactly written by introducing the finite-difference operator $\lO$ 
\begin{align}\label{Baxter1}
    &\mathcal{O} = \frac{1}{Q^+_{\theta}Q^-_{\theta}}\left[\chi_{+3}{Q_{{\theta}}^{-}}D^{3}{Q_{\Bar{\theta}}^{+}}
    -
    \tau_+\, D^{+}
    +\tau_-\, D^{-}    
    -\chi_{-3}{Q^{+}_{\theta}}D^{-3}{Q^{-}_{\bar\theta}}\right]\frac{1}{Q^{+}_{\Bar{\theta}}Q^{-}_{\Bar{\theta}}}\,.
\end{align}
Let us unpack the notation. $D^\pm$ are shift operators, which can act on functions either on the left or the right, by
\begin{equation}
    \overset{\rightarrow}{D^\pm}\, f  = f^\pm ,\quad f\,\overset{\leftarrow}{D^\pm} = f^\mp\,.
\end{equation}
$\tau_\pm(u)$ are polynomials of degree $2L$, and are the eigenvalues of transfer matrices in fundamental and anti-fundamental representations $\hat \tau_\pm(u)$, which contain all $4L$ independent integrals of motion $\hat I_{\pm,n}$ of the system 
\begin{equation}\label{defI}
    \tau_\pm (u) = \chi_\pm \left(u\pm\frac{i}{4} \right)^{2L}+\displaystyle \sum_{n=0}^{2L-1}\left(u\pm\frac{i}{4} \right)^{n}I_{\pm,n}
\end{equation}
where $I_{\pm,n}$ are the eigenvalues of $\hat{I}_{\pm,n}$. Note that the transfer matrices are not new objects, but are themselves fixed in terms of the Q-functions. For example:
\begin{equation}\label{eq:TauFromQsu3}
\begin{split}\
& \tau_+(u)=Q_\theta^+ Q_{\bar\theta}^{[2]}\frac{Q_1^{[-2]}}{Q_1} + \chi_{-3} Q_\theta^- Q_{\bar\theta}^{[2]} \frac{Q_1^{[2]}} {Q_1}\frac{Q^{3\, -}}{Q^{3\,+}}+Q_\theta^- Q_{\bar\theta}\frac{Q^{3\, [3]}}{Q^{3\,+}}\;.
\end{split}
\end{equation}

Finally $\chi_A$ are characters of the twist matrix, which can be defined in terms of the twist matrix eigenvalues $\lambda_1,\;\lambda_2,\;\lambda_3$ as
\beq
\chi_{+3}=1\;,\quad
    \chi_+ = \lambda_1 + \lambda_2 +\lambda_3\;,\quad \chi_- = \lambda_1 \lambda_2 + \lambda_1 \lambda_3 +\lambda_2 \lambda_3 \;,\quad \chi_{-3} = \lambda_1 \lambda_2 \lambda_3\;.
\eeq

Having made these definitions, the Baxter equation and its dual can be both compactly written as 
\begin{equation}\label{eqn:2Baxters}
   Q_a \overset{\leftarrow}{\cal O}= 0\;, \quad \overset{\rightarrow}{\cal O} Q^a = 0\;,
\end{equation}
where the arrows over $\lO$ denote in which direction the shift operators $D^\pm$ act. 
We see that it would be useful to absorb  the factors in denominator of \eq{Baxter1} into the Q-functions and define
\beq\la{notation}
\bQ_a\equiv \frac{Q_a}{Q_\theta^+ Q_\theta^-}\;\;,\;\;\bQ^a\equiv \frac{Q^a}{Q_{\btheta}^+ Q_{\btheta}^-}\;.
\eeq
We will make use of this notation in the next section.

\section{Crash Course in Functional Separation of Variables}\la{sec:crash_course_fsov}
The Baxter equations encode the complete information of the spectrum of conserved quantities. The philosophy of Functional Separation of Variables (FSoV) is that the Baxter equation contains sufficient information to compute \textit{all} observables in the system, not just the spectrum of conserved quantities. In this section we will explain how FSoV allows us to compute the scalar product $\braket{\Psi_{B}}{\Psi_{A}}$, where $\langle \Psi_B|$ and $|\Psi_A\rangle$ are left and right transfer matrix eigenstates, respectively. Due to our set-up with an alternating spin chain we expand the construction presented in \cite{Gromov:2022waj} to this case. This section will naturally set the stage for the computation of $\braket{W}{\Psi_{A}}$ to be presented in Section~\ref{sec:WPsiSection}.

\subsection{Basics of Functional Separation of Variables}

\paragraph{Measure and adjointness.}
The starting point for FSoV is the definition of the family of linear forms
\begin{equation}
    \bl f \br = \oint \frac{du}{2\pi \ii}\rho(u)\, f(u) 
\end{equation}
depending on the measure $\rho$, which is any entire $i$-periodic function. The integration contour is assumed to encircle all poles of the function $f$. 

With this definition, we have the following adjointness property of the Baxter finite-difference operator
\beq\la{selfadj}
\bl f \overset{\rightarrow}{\mathcal{O}} g \br = \bl \Big(f \overset{\leftarrow}{\mathcal{O}} g\Big)^+ \br\;.
\eeq

In this paper we will be focusing on functions $f$ which have poles at $\theta_a+ik+\tfrac{i}{2}$ and at $-\theta_a+i k$ for some integers $k$. For such functions, any two different choices of the functions $\rho$ whose difference is a periodic function with zeroes at all $\theta_\beta+i/2$ and $-\theta_\beta$, $\beta=1,\dots,L$ will produce the same result for the integral. Hence, there are only $2L$ linearly independent linear forms. 

We will denote a basis of these linear forms by
\beq\la{rhoalpha1}
\rho_\alpha(u)=\frac{1}{{\cal R}_\alpha}\prod_{\beta\neq \alpha}^L (1+e^{2\pi (u-\theta_\beta)})
\prod_{\beta=1}^L (1-e^{-2\pi (u+\theta_\beta)})\;\;,\;\;\rho_{\bar\alpha}(u)=\rho_{\alpha}(-u+\tfrac i2)\;,
\eeq
where the constant ${\cal R}_\alpha$ is chosen so that $\rho_{\alpha}(\theta_\alpha+\tfrac i2)=1$.
We will denote the corresponding linear forms by 
\begin{equation}\la{brdef}
    \bl f \br_\alpha = \oint \frac{du}{2\pi \ii}\rho_\alpha(u)\, f(u)\,,\quad \bl f \br_{\bar\alpha} = \oint \frac{du}{2\pi \ii}\rho_{\bar\alpha}(u)\, f(u)\,,
\end{equation}
for $\alpha=1,\dots,L$. 
These linear forms will play a key role in the remainder of the text. 

\subsection{Computing the Scalar Product}\label{sec:ScalarProductFSoV}
The starting point is the following trivial identity for two states $A$ and $B$
\begin{equation}\label{eqn:trivdiff1}
\begin{split}
   &  \bl Q_b^A\Big(\overset{\rightarrow}{\lO_A} - \overset{\rightarrow}{\lO_B} \Big) Q^a_B \br_{\alpha} = 0\,
\end{split}
\end{equation}
valid for any choice of $a,\;b=1,\;2,\;3$, which follows from the Baxter equations \eqref{eqn:2Baxters} and the adjointness property \eqref{selfadj}. We will only consider the case $b=1$ and $a=1,\;3$. 

Using \eq{selfadj} we now expand the difference of operators $\lO_{A,B}$ into their constituent parts. To accommodate the various factors of $Q_\theta$ and $Q_{\bar\theta}$ we will be using the notation \eq{notation},
which will render many expressions more compact. In this notation, the expansion of the difference operators in \eqref{eqn:trivdiff1} yields
\begin{equation}\label{eqn:splinsys}
\begin{split}
   & \displaystyle \sum_{p=1}^{2L}\bl \bQ^A_1 \left(u+\tfrac{i}{4}\right)^{p-1} \bQ_B^{a,+}  \br_{\alpha} I^{AB}_{+,p-1} - \bl \bQ_1^A \left(u-\tfrac{i}{4}\right)^{p-1} \bQ^{a,-}_B  \br_{\alpha} I^{AB}_{-,p-1} = 0\;,
\end{split}
\end{equation}
where we have denoted the differences of the integrals of motion $I_{\pm,n}^{AB}:=I_{\pm,n}^A - I_{\pm,n}^B$. 
Interchanging $a$ and 1 in \eq{eqn:trivdiff1} and taking a different linear form defined with $\bar\alpha$ this time we get
\begin{equation}\label{eqn:splinsys20}
\begin{split}
   & \displaystyle \sum_{p=1}^{2L}\bl \bQ^A_a \left(u+\tfrac{i}{4}\right)^{p-1} \bQ_B^{1,+}  \br_{\bar\alpha} I^{AB}_{+,p-1} - \bl \bQ_a^A \left(u-\tfrac{i}{4}\right)^{p-1} \bQ^{1,-}_B  \br_{\bar\alpha} I^{AB}_{-,p-1} = 0\;.
\end{split}
\end{equation}
Next we notice that the equations \eqref{eqn:splinsys}  and \eqref{eqn:splinsys20} together constitute a system of $4L$  equations (for different $\alpha$ and $a$) for the $4L$ unknowns $I^{AB}_{\pm,p}$. For the conserved charges to have non-degenerate spectrum, at least one of the differences $I^{AB}_{\pm,p}$ must be non-zero, otherwise we would have two distinct states with the same values of all conserved charges. As a result, for $A\neq B$, the determinant of the coefficient matrix must vanish, providing a natural notion of a scalar product on the Hilbert space of the spin chain. Indeed, since in our set-up the transfer matrices have non-degenerate spectrum, a left and right eigenstate $|\Psi_A\rangle$ and $\langle \Psi_B|$ are orthogonal for $A\neq B$, a feature manifest in the determinant of the coefficient matrix. Since transfer matrix eigenstates are uniquely characterised by Q-functions $\bQ_a^A$ and $\bQ^a_B$ we should then have
\begin{equation}\label{eqn:spdet}
    \langle \Psi_B|\Psi_A\rangle = \frac{1}{\cal N}\displaystyle\det_{(a,\alpha),(s,p)} \left(
    \begin{array}{c}
        \bl \bQ_1^A\; D^{\tfrac{s}2} \;u^{p-1}\; D^{\tfrac{s}2}\; \bQ_B^a\br_\alpha   \\
        \bl \bQ_a^A\; D^{\tfrac{s}2} \;u^{p-1}\; D^{\tfrac{s}2}\;\bQ^1_B\br_{\bar\alpha}  
    \end{array}
    \right)\;,
\end{equation}
where the $4L\times 4L$ matrix is parameterised by the following indices
$a=1,3$, $s=\pm $, $\alpha=1,\dots,L$
and $p=1,\dots,2L$. Thus for example for $L=1$ 
\begin{equation}
\begin{split}
    &\langle \Psi_B|\Psi_A\rangle = \\
    &\displaystyle\det \begin{pmatrix}
        \bl \bQ_1^{A} D \bQ^{1}_{B} \br_{1} & \bl \bQ_1^{A} D^{\frac{1}{2}}uD^{\frac{1}{2}} \bQ^{1}_{B} \br_{1} & \bl \bQ_1^{A} D^{-1} \bQ^{1}_{B} \br_{1} & \bl \bQ_1^{A} D^{-\frac{1}{2}}uD^{-\frac{1}{2}} \bQ^{1}_{B} \br_{1} \\
        \bl \bQ_1^{A} D \bQ^{1}_{B} \br_{\bar{1}} & \bl \bQ_1^{A} D^{\frac{1}{2}}uD^{\frac{1}{2}} \bQ^{1}_{B} \br_{\bar{1}} & \bl \bQ_1^{A} D^{-1} \bQ^{1}_{B} \br_{\bar{1}} & \bl \bQ_1^{A} D^{-\frac{1}{2}}uD^{-\frac{1}{2}} \bQ^{1}_{B} \br_{\bar{1}}\\
        \bl \bQ_1^{A} D \bQ^{3}_{B} \br_{1} & \bl \bQ_1^{A} D^{\frac{1}{2}}uD^{\frac{1}{2}} \bQ^{3}_{B} \br_{1} & \bl \bQ_1^{A} D^{-1} \bQ^{3}_{B} \br_{1} & \bl \bQ_1^{A} D^{-\frac{1}{2}}uD^{-\frac{1}{2}} \bQ^{3}_{B} \br_{1}\\
        \bl \bQ_3^{A} D \bQ^{1}_{B} \br_{\bar{1}} & \bl \bQ_3^{A} D^{\frac{1}{2}}uD^{\frac{1}{2}} \bQ^{1}_{B} \br_{\bar{1}} & \bl \bQ_3^{A} D^{-1} \bQ^{1}_{B} \br_{\bar{1}} & \bl \bQ_3^{A} D^{-\frac{1}{2}}uD^{-\frac{1}{2}} \bQ^{1}_{B} \br_{\bar{1}}
    \end{pmatrix}\;.
\end{split}
\end{equation}

What is less trivial to see is that it is also non-zero for $A=B$. Of course, for this equation to hold literally we must specify the normalisation of the transfer matrix eigenstates $\langle \Psi_B|$ and $|\Psi_A\rangle$. We will do this in the next subsection and obtain a clear interpretation of \eqref{eqn:spdet} as a scalar product in a separation of variables (SoV) basis.

\subsection{Linking to SoV Bases.}\label{linkSoV}

\paragraph{SoV basis.} In this section we assume that that there exist a basis $\langle \svx|$ such that the eigenstates of the Hamiltonian factorise into a product, i.e. a separation of variables (SoV) basis. In Appendix \ref{app:sov} we will describe how to build such a basis explicitly -- here we will simply state its properties. 

The basis $\langle \svx|$ is parameterised by $4L$ numbers $\svx_{\alpha,a}$ and $\bar{\svx}_{\alpha,a}$ given by
\begin{equation}\la{eq:xinn}    \svx_{\alpha,a} = +\theta_\alpha + i\, n_{\alpha,a}-\frac{i}{2}\;,\quad \bar{\svx}_{\alpha,a} = -{\theta}_\alpha + i\,(\bn_{\alpha,a}+1-a)\;,
\end{equation}
where $n_{\alpha,a}$ and $\bar{n}_{\alpha,a}$ are non-negative integers subject to the conditions
\begin{equation}\label{eqn:ninequal}
    1 \geq n_{\alpha,1} \geq n_{\alpha,2} \geq 0\;,\quad 1 \geq \bn_{\alpha,1} \geq \bn_{\alpha,2} \geq 0\;.
\end{equation}

The defining feature of an SoV basis in integrable spin chains is that the wave functions $\Psi(\svx):=\langle \svx|\Psi\rangle$ of the transfer matrix eigenstates factorise into a product of Q-functions. The precise form of this factorisation depends on the representation in the physical Hilbert space -- see \cite{Ryan:2018fyo,Ryan:2020rfk} where this factorisation was determined for $\mathfrak{su}(N)$ spin chains in any finite-dimensional representation. For the alternating fundamental--anti-fundamental chain considered here we have
\begin{equation}\label{eqn:wavepropto}
    \langle \svx|\Psi\rangle\, \propto \,\displaystyle\prod_{\alpha=1}^L 
Q_1(\svx_{\alpha,1})Q_1(\svx_{\alpha,2})\left[Q_1(\bar{\svx}_{\alpha,1})Q_3(\bar{\svx}_{\alpha,2})-
Q_1(\bar{\svx}_{\alpha,2})Q_3(\bar{\svx}_{\alpha,1})\right]\,.
\end{equation}
In previous SoV studies \cite{Gromov:2016itr,Ryan:2018fyo,Gromov:2019wmz,Ryan:2020rfk,Gromov:2020fwh,Gromov:2022waj}, the $\Psi$-independent proportionality factor was simply set to be $1$. In the current set-up we have found it advantageous to consider a different normalisation involving various $Q_\theta$ factors and in particular replacing $Q$ with  $\bQ$. However, we must be careful with this replacement in \eqref{eqn:wavepropto} since $\bQ(u)$ has poles at precisely $\svx_{\alpha,a}=\theta_\alpha\pm \frac{i}{2}$. To account for this, we will express the overlap in terms of the residues of $\bQ_1$ at these poles or, equivalently, as a multiple contour integral
\begin{equation}\label{eqn:xwvfn}
    \langle \svx|\Psi\rangle =\displaystyle \prod_{\alpha=1}^L
    \oint_{\mathsf{x}_{\alpha,a}} \frac{du_1}{2\pi i}\frac{du_2}{2\pi i}\rho_\alpha(u_1)\rho_\alpha(u_2)    
\bQ_1(u_1)\bQ_1(u_2)\left[\bQ_1(\bar{\svx}_{\alpha,1})\bQ_3(\bar{\svx}_{\alpha,2})-
\bQ_1(\bar{\svx}_{\alpha,2})\bQ_3(\bar{\svx}_{\alpha,1})\right]
\end{equation}
where the contour of integration goes around the pole at $u_a= \svx_{\alpha,a}$.

\paragraph{Dual SoV basis.}

In addition to the basis $\langle \svx|$, there is a dual SoV basis $|\svy \rangle$~\cite{Gromov:2019wmz} with the property that it factorises the \textit{left} transfer matrix eigenstates $\langle \Psi|\svy\rangle$ into a product of Q-functions. 

The basis $|\svy\rangle$ is again parameterised by $4L$ numbers $\svy_{\alpha,a}$ and $\bar{\svy}_{\alpha,a}$ with 
\begin{equation}
 \la{ym}   {\svy}_{\alpha,a} = +{\theta}_\alpha + i\,({m}_{\alpha,a}+1-a)\;,\quad \bar{\svy}_{\alpha,a} = -\theta_\alpha + i\, \bm_{\alpha,a}-\frac{i}{2}\;,
\end{equation}
subject to the selection rules
\begin{equation}\label{eqn:minequal}
    1 \geq m_{\alpha,1} \geq m_{\alpha,2} \geq 0\;,\quad 1 \geq {\bm}_{\alpha,1} \geq {\bm}_{\alpha,2} \geq 0\;.
\end{equation}

In the same manner as with $\langle \svx|$, we normalise the left wave functions as 
\begin{equation}\label{eqn:ywvfn}
    \langle \Psi| \svy\rangle =\displaystyle \prod_{\alpha=1}^L
    \oint_{\bar\svy_{\alpha,a}} \frac{du_1}{2\pi i}\frac{du_2}{2\pi i}\rho_{\bar\alpha}^+(u_1)\rho_{\bar\alpha}^+(u_2)    
\bQ^1(u_1)\bQ^1(u_2)\left[\bQ^1({\svy}_{\alpha,1})\bQ^3({\svy}_{\alpha,2})-
\bQ^1({\svy}_{\alpha,2})\bQ^3({\svy}_{\alpha,1})\right]\,,
\end{equation}
here the integral in $u_a$ goes around the pole $\bar\svy_{\alpha,a}$ only.

\paragraph{Expansion into SoV bases.}
We now explain how the Functional Separation of Variables is related to the SoV bases $\langle \svx|$ and $|\svy\rangle$ described in the preceding paragraphs.  

We begin by considering the overlap $\langle \Psi_B|\Psi_A\rangle$ of two transfer matrix eigenstates $\langle \Psi_B|$ and $|\Psi_A\rangle$. We can compute this overlap by performing two resolutions of the identity 
\begin{equation}
    1 = \displaystyle\sum_{\svx} |\svx\rangle \langle \svx|\;,\quad 1 = \displaystyle\sum_{\svy} |\svy\rangle \langle \svy|
\end{equation}
where $|\svx\rangle$ and $\langle \svy|$ are canonically conjugate dual vectors to $\langle \svx|$ and $|\svy\rangle$ respectively:
\begin{equation}\la{idres}
    \langle \svx|\svx'\rangle = \delta_{\svx \svx'}\;,\quad \langle \svy|\svy'\rangle = \delta_{\svy \svy'}\;.
\end{equation}
The computation of the overlap then amounts to computing 
\begin{equation}\label{SoVsp}
  \langle \Psi_B|\Psi_A\rangle=  \displaystyle \sum_{\svx,\svy} \Psi_B(\svy)\lM_{\svy,\svx}\Psi_A(\svx) \;,
\end{equation}
where we have introduced the SoV measure $\lM_{\svy,\svx}:=\langle \svy|\svx\rangle$. 

The explicit computation of the measure $\lM_{\svy,\svx}$ starting from the microscopic description of the spin chain is extremely non-trivial, see \cite{Maillet:2020ykb} for an implicit characterisation of the measure using this approach. One of the powerful features of the Functional Separation of Variables is that it allows us to compute the measure explicitly and efficiently~\cite{Gromov:2020fwh}. Indeed, by expanding the determinant \eqref{eqn:spdet} and performing the integration by residues we immediately see that it has the form \eqref{SoVsp}! 

In order to fix the normalisation factor $\cal N$ in \eqref{eqn:spdet} to ensure that \eqref{eqn:spdet} matches \eqref{SoVsp} we have to require that ${\cal M}_{0,0}=1$. For this, one can extract the relevant term from the expansion in the sum over residues giving
\beq\la{calN}
{\cal N}=\Delta\(\{+\theta_\alpha-\tfrac{i}{4}\}\cup\{-\theta_\alpha-\tfrac{3i}{4}\}\)\;\;\times\;\;
\Delta\(\{-\theta_\alpha-\tfrac{i}{4}\}\cup\{+\theta_\alpha-\tfrac{3i}{4}\}\)
\eeq
where $\Delta$ is a Vandermonde determinant\footnote{Given an ordered set $A$ with elements $A_1,A_2,\dots$ we define $\Delta(A) =\prod_{i<j}(A_i-A_j)$.}. In Appendix~\ref{app:sigmification} we explicitly computed all ${\cal M}_{\svy,\svx}$ starting from \eq{eqn:spdet}. What we found is the following compact expression for the determinant in the r.h.s. of \eq{eqn:spdet}
\beqa\la{detnorminSoV}
{\rm det}=&&
\sum_{\svx,\;\svy}
\Psi_B(\svy)
\Psi_A(\svx)
\sum_\sigma{\rm sig}_\sigma\;
\delta\(
{{\svy} - {\svx}+i\sigma+\tfrac{3i}{2}}
\)
\prod_{a=1,2}\Delta\(\frac{{\svx}_{\sigma^{-1}(a)}+{\svy}_{\sigma^{-1}(a)}}{2}\)\;,
\eeqa
where the sum goes over all SoV states i.e. all $n,\;\bar n$ and $m,\;\bar m$, satisfying  \eq{eqn:ninequal} and \eq{eqn:minequal}. In addition we have to sum over permutations $\sigma$ of the $4L$ numbers $1,2,1,2,\dots$ such that $\sigma_{p,a}$ with $p=1,\dots,2L$ and $a=1,2$ is either $1$ or $2$. Finally, $\svx_{\sigma^{-1}(a)}+\svy_{\sigma^{-1}(a)}$ gives the set of all numbers $\svx_{p,a}+\svy_{p,a}$ for which $\sigma_{p,a}=a$, where the index $p$ combines $\alpha$ and $\bar\alpha$ in a convenient way i.e. 
${p\;{\rm mod}\;L}=\alpha\;{\rm if}\;p\leq L\;{\rm else}\;{\bar\alpha}$.

\section{Discrete Symmetry and Integrable Boundary State Overlaps}\label{sec:WPsiSection}

\subsection{Reflection Symmetry}

For our twisted spin chain solutions of the Baxter TQ equations \eqref{eqn:2Baxters} and the QQ-relations \eqref{eqn:QQrelns} are in one-to-one correspondence with transfer matrix eigenstates. These equations have the following $\mathbb{Z}_2$ symmetry, which we refer to as \textit{reflection}:
\beqa\label{symmetry}
\nonumber Q_a(u) \leftrightarrow Q^a(-u)\;, \quad \bQ_a(u) \leftrightarrow \bQ^a(-u)\;, \quad \tau_{\pm}(u) \leftrightarrow \tau_{\pm}(-u\mp i/2)\;.
\eeqa

Thanks to the non-degeneracy of the spectrum this symmetry must map one state to another. More precisely, in general for a state labelled $A$, there is another state labelled $\bar{A}$ such that 
\beq\la{Qrefl}
Q^A_a(u)= (-1)^{M_a}  Q^{a}_{\bar{A}}(-u)\;,\quad
\bQ^A_a(u)= (-1)^{M_a}  \bQ^{a}_{\bar{A}}(-u)\;.
\eeq

\paragraph{State selection rules.}

There are two types of states -- those which are symmetric under reflection (in which case $\bar{A}=A$), or states not invariant under this symmetry and hence $\bar{A}\neq A$. One can count that there are exactly $3^L$ symmetric states as we argue in Section \ref{genboundaries}.

Let us examine the consequences on the integrals of motion for two states related by reflection symmetry. As described in the previous section, for these states the transfer matrix eigenvalues must satisfy
\begin{equation}\label{Tsymmetry}
     \tau_+^A(u) = \tau^{\bar A}_+(-u-i/2)\;,\quad  \tau^A_-(u) = \tau^{\bar A}_-(-u+i/2)\;.
\end{equation}
We can expand these functions into a complete basis of integrals of motion $I_{\pm,n}$  as 
\begin{equation}
     \tau_\pm (u) = \chi_\pm \left(u\pm\frac{i}{4} \right)^{2L}+\displaystyle \sum_{n=0}^{2L-1}\left(u\pm\frac{i}{4} \right)^{n}I_{\pm,n}\,.
\end{equation}
From \eqref{Tsymmetry} we see that for such states the integrals of motion must be related as 
\begin{equation}\label{Iselection}
    I_{\pm,n}^A = (-1)^n I_{\pm,n}^{\bar A}\;\quad n=0,1,\dots,2L-1\;.
\end{equation}
Specialising to a state which is reflection symmetric, i.e. $\bar{A}=A$, we conclude that $I_{\pm,\text{odd}}=0$. 

\paragraph{Q-function selection rules.} The degrees of the polynomial parts of the Q-functions are constrained for states related by reflection. Denoting by $M_a$ and $M^a$ the degrees of $Q_a^A$ and $Q^a_A$, and $\bar{M}_a$ and $\bar{M}^a$ the degrees of $Q_a^{\bar A}$ and $Q^a_{\bar A}$ respectively, the reflection property \eq{symmetry} then implies that $\bar{M}^a = M_a$ and $\bar{M}_a = M^a$. 

Furthermore, for an invariant state where $\bar{A} = A$, we thus have $M^a = M_a$, which due to \eq{wr2} and \eq{wr3} implies $M_a = M^a = L$, $a=1,2,3$.

\subsection{Block-Diagonal Form of the Scalar Product}
We now investigate the consequences of reflection symmetry on the scalar product of transfer matrix eigenstates. We demonstrate that the matrix in the scalar product $\langle \Psi_B|\Psi_A\rangle$ given by the determinant \eq{eqn:spdet} can be reduced to a block-diagonal form in the case when $B$ is the state obtained from $A$ by action of reflection i.e. satisfies \eq{symmetry}. Here we follow the same logic as in \cite{Caetano:2020dyp,Cavaglia:2021mft}. 

For the readers convenience we will repeat the scalar product here in the case $B=\bar{A}$:
\begin{equation}
     \langle \Psi_{\bar A}|\Psi_A\rangle = \frac{1}{\cal N}\displaystyle\det_{(a,\alpha),(s,p)} \left(
    \begin{array}{c}
        \bl \bQ_1^A\; D^{\tfrac{s}2} \;u^{p-1}\; D^{\tfrac{s}2}\; \bQ_{\bar A}^a\br_\alpha   \\
        \bl \bQ_a^A\; D^{\tfrac{s}2} \;u^{p-1}\; D^{\tfrac{s}2}\;\bQ^1_{\bar A}\br_{\bar\alpha}  
    \end{array}
    \right)\;.
\end{equation}
We will now perform various transformations on the determinant leaving it invariant but reducing the matrix to a simpler form. In the bottom $2L$ rows we change the integration variable from $u\to -u$ and move the shift operator to act on $\bQ_{a}$. More precisely we use
\beq\la{qqqq}
\bl \bQ_a^A\; D^{\tfrac{s}2} \;u^{p-1}\; D^{\tfrac{s}2}\;\bQ^1_{\bar A}\br_{\bar\alpha} 
=
(-1)^{p}\bl \bQ^1_{\bar A}(-u)\; D^{\tfrac{s}2} \;u^{p-1}\; D^{\tfrac{s}2}\;\bQ_a^A(-u)\br_{\alpha} 
\eeq
where we also used the relation \eq{rhoalpha1} to convert $\alpha$ to $\bar\alpha$. Next using the reflection symmetry  of the Q-functions \eq{Qrefl} in the r.h.s. of \eq{qqqq} we can convert $-u$ arguments back to get
\beq
\bl \bQ_a^A\; D^{\tfrac{s}2} \;u^{p-1}\; D^{\tfrac{s}2}\;\bQ^1_{\bar{A}}\br_{\bar\alpha} 
=
(-1)^{p}\bl \bQ_1^A\; D^{\tfrac{s}2} \;u^{p-1}\; D^{\tfrac{s}2}\;\bQ^a_{\bar{A}}\br_{\alpha} 
\eeq
i.e. up to a sign we get the same expression as in the first $2L$ rows. Adding and subtracting the corresponding rows
and rearranging the columns we get two $2L\times 2L$ blocks on the diagonal 
\begin{equation}
    \langle \Psi_{\bar A}|\Psi_A\rangle = \frac{4^{L}}{{\cal N}}\displaystyle\det_{(a,\alpha),(s,\beta)} \left(
    \begin{array}{cc}
        \bl \bQ_1^A\; D^{\tfrac{s}2} \;u^{2\beta-1}\; D^{\tfrac{s}2}\; \bQ^a_{\bar A}\br_\alpha &      0    \\
        0 &
        \bl \bQ_1^A\; D^{\tfrac{s}2} \;u^{2\beta-2}\; D^{\tfrac{s}2}\; \bQ^a_{\bar A}\br_\alpha
    \end{array}
    \right)\;,
\end{equation}
where $\alpha,\beta=1,\dots,L$, $a=1,3$ and $s=\pm$. 

We denote the determinant of the upper block by $\det_+$ and the determinant of the lower block by $\det_-$. When the state is not invariant under reflection, and so $\bar{A}$ and $A$ correspond to two distinct states, the overlap $\langle \Psi_{\bar A}|\Psi_A\rangle$ must vanish. Hence, one of the determinants $\det_+$ or $\det_-$ must also vanish. In the next section we prove that $\det_+$ is the one which vanishes, and also show that $\det_+$ is precisely the overlap of $|\Psi_A\rangle$ with a special state $\langle W|$, annihilated by all odd integrals of motion and which is a singlet under the $\mathfrak{su}(3)$ symmetry, known in the literature as a {\it dimer} state (see e.g.~\cite{pozsgay2014overlaps}). 

\subsection{The Boundary State $\bra{W}$}\la{sec:bs}
We now define a so-called dimer integrable boundary state $\langle W|$, following \cite{Ghoshal:1993tm,Piroli:2017sei} -- a state annihilated by all odd charges:
\beq\la{Wdef}
\langle W|\hat{I}_{\pm,2\beta-1} = 0\;\;,\;\;\beta=1,\dots,L
\eeq
or equivalently in terms of transfer matrices 
\begin{equation}\label{Wdeftau}
    \langle W|\hat{\tau}_+(-u)= \langle W|\hat{\tau}_+(u-i/2),\quad \langle W|\hat{\tau}_-(-u)= \langle W|\hat{\tau}_-(u+i/2)\,.
\end{equation}
This definition does not single out $\langle W|$ uniquely. There are many possible integrable boundary states, for example Matrix Product States, two-site states, etc. These are usually defined in the microscopic description of the spin chain, e.g. as a vector with specific components in the standard basis. We will show that there is a distinguished integrable boundary state naturally encoded in the Functional Separation of Variables formalism. This state is uniquely distinguished by the fact that it is a singlet under the global symmetry algebra, as discussed in Appendix \ref{app:basis_integrable_states}. From now on when we write $\langle W|$ it will refer exclusively to this state. We now show how to construct this state in the FSoV formalism. 

\subsection{FSoV for Integrable Boundary State Overlaps}\label{sec:FSoVOverlap}

The starting point is \eqref{eqn:trivdiff1}, which we rewrite here for convenience, 
\begin{equation}\label{eqn:trivdiff2}
\bl Q_b^A(u)(\overset{\rightarrow}{\lO_A} - \overset{\rightarrow}{\lO_{B}} ) Q_{B}^a(u) \br_{\alpha} = 0\, 
\end{equation}
and we specialise to the case where the state $A$ is a generic state and the state $B = \bar{A}$ is obtained from $A$ by the reflection operation~\eq{symmetry}. Due to the relation \eqref{Iselection} only the odd charges survive in \eq{eqn:trivdiff2}
giving
\begin{equation}\label{eqn:splinsys2}
\begin{split}
   & \displaystyle \sum_{\beta=1}^{L}\sum_{s=\pm 1}s\;\bl \bQ^A_1(u) D^{\frac{s}{2}} u^{2\beta-1} D^{\frac{s}{2}}\bQ^A_a(-u)  \br_{\alpha}\; I_{s,2\beta-1} = 0\;.
\end{split}
\end{equation}
where also used \eq{Qrefl} to lower the indices. We will omit the state label $A$ since it is now the same for all Q-functions. Again considering \eq{eqn:splinsys2} as a system of $2L$ equations (for $a=1,3$ and $\alpha=1,\dots,L$) for $2L$ odd charges as we did with the scalar product, we conclude that for states which are not reflection invariant, i.e. $\bar{A}\neq A$ and so at least some odd charges must be non-zero, the below $2L\times 2L$ determinant must vanish
\begin{equation}\label{eqn:spdet2}
    {\rm det}_+=\displaystyle\det_{(a,\alpha),(s,\beta)}         \bl \bQ_1(u)\; D^{\tfrac{s}2} \;u^{2\beta-1}\; D^{\tfrac{s}2}\; \bQ_a(-u)\br_{\alpha} \;.
\end{equation}
Again, less obvious is that this determinant is non-zero for the symmetric states. It will become clear once we interpret this determinant as an overlap $\langle W|\Psi\rangle$ in separated variables.

As we did with the scalar product, by expanding out the determinant and computing all integrals by residues, we can write $\det_+$ in the following form
\beq\la{detdec}
{\rm det}_+ =\sum_{{\svx}} {\cal W}_{\svx} \Psi(\mathsf{x})
\eeq
where the sum goes over the complete basis of SoV states. In Appendix~\ref{app:Wx} we perform this expansion and compute the coefficients ${\cal W}_{\svx}$ explicitly. Crucially we find that $\cal W_{\svx}$ are independent of the choice of $Q$-functions in $\det_+$. Furthermore, an important property we will make use of later is that all ${\cal W}_{\svx}$ are independent of the twist parameters $\lambda_1,\lambda_2,\lambda_3$.

Now we can define the state $\langle W|$ mentioned in the previous subsection: it is defined by these components in the SoV basis
\beq\la{Wrevert}
\langle W| \equiv  \sum_{{\bf x}} {\cal W}_{\svx} \langle{\svx}|
\eeq
and it is immediate that its overlap with $|\Psi\rangle$ is given by
\begin{equation}\label{WPsioverlap}
    \langle W|\Psi\rangle = \displaystyle\det_{(a,\alpha),(s,\beta)}         \bl \bQ_1(u)\; D^{\tfrac{s}2} \;u^{2\beta-1}\; D^{\tfrac{s}2}\; \bQ_a(-u)\br_{\alpha} \;.
\end{equation}
Let us note that, firstly, $\langle W|$ defined by \eqref{Wrevert} is indeed a non-trivial state as at least some ${\cal W}_{\bf x}$ are non-zero as demonstrated in Appendix~\ref{app:Wx}. Secondly, as we have shown above the r.h.s. in \eqref{WPsioverlap} must vanish for states $|\Psi\rangle$ which are not reflection invariant and hence $\langle W|$ is orthogonal to all such states. In particular, since the set of all integral of motion eigenstates form a basis, this implies that $\langle W|$ must be annihilated by all odd charges, i.e. $\langle W|$ defines an integrable boundary state \eqref{Wdef}. As, in addition, it does not depend on the twist it is annihilated by the odd charges of the spin chain with arbitrary values of the twists parameters $\lambda_a$. As the highest odd charges $I_{\pm,2L-1}$ are built out of a combination of $\mathfrak{su}(3)$ generators with coefficients dependent on $\lambda_a$ these must annihilate it. In appendix \ref{app:properties_w} we show that this is sufficient to show that $\langle W|$ must be a singlet state, which singles it out to be the dimer boundary state.

\subsection{Normalisation of the Boundary State}
We can also define $|W\rangle$ and determine the 
overlap $\langle \Psi| W\rangle$ by rewriting ${\rm det}_+$ as follows using the symmetry of the state
\begin{equation}\label{eqn:spdet2Norm}
    {\rm det}_+=\displaystyle\det_{(a,\alpha),(s,\beta)}         \bl \bQ^1(-u)\; D^{\tfrac{s}2} \;u^{2\beta-1}\; D^{\tfrac{s}2}\; \bQ^a(u)\br_{\alpha} \;.
\end{equation}
Where the $Q$-function are those of the symmetry-transformed state $\bar A$. But since we know that the determinant must be zero for the states for which $A\neq \bar A$ we have in general $\langle W|\Psi\rangle = \langle \Psi| W\rangle$. In the form \eq{eqn:spdet2Norm} we can then define $\langle \Psi| W\rangle \equiv {\rm det}_+$ as it allows for the decomposition
\beq\label{detWy}
{\rm det}_+ = \sum_{\svy}{\cal W}_{\svy}\langle\Psi |{\svy}\rangle
\;.
\eeq
Where we can see that the SoV representation of the left and right $W$ states
$W_\svx\equiv\langle W|{\svx}\rangle$
and $W_\svy\equiv\langle {\svy}|W\rangle$ are related by a simple replacement of the SoV indexes
\beq
n_{\alpha, a}\to \bar m_{\alpha,3-a}\;\;,\;\;
\bar n_{\alpha, a}\to m_{\alpha,3-a}\;.
\eeq
Finally, we can compute the overlap $\langle W|W\rangle$ by inserting two resolutions of identity like in \eq{idres} to get
\beq\label{WWover}
\langle W|W\rangle =
\sum_{{\svx},{\svy}}\langle W|{\svx}\rangle {\cal M}^{-1}_{{\svy},{\svx}}
\langle {\svy}|W\rangle\;.
\eeq
As we have an efficient way of evaluating all the $3$ quantities: the measure and the overlaps $\langle {\svy}|W\rangle$ and $\langle W|{\svx}\rangle$ we found the following simpler relation to hold for $L=1,2,3$
\beq\la{Wnorm}
\langle W|W\rangle =\left(\frac{3}{4}\right)^LQ_\theta(+\tfrac{i}{4})Q_\theta(-\tfrac{i}{4})\;{\cal N} 
\eeq
which we expect to hold for general $L$. Here ${\cal N}$ is defined in \eq{calN}. The equation \eq{Wnorm} thus fixes the state $\langle W|$.

\paragraph{Normalisation independent overlap.}

The overlaps we have computed depend on the normalisations of the various states involved. We will now form a normalisation-independent combination. Combining \eqref{eqn:spdet},\eqref{detdec}, \eqref{detWy} and\eqref{WWover} we obtain 
\begin{equation}\label{normalisationindep}
    \frac{\langle W|\Psi\rangle\langle \Psi|W\rangle}{\langle W| W\rangle \langle \Psi|\Psi\rangle}= \frac{1}{3^{L}Q_\theta(\tfrac{i}{4})Q_\theta(-\tfrac{i}{4})} \frac{\det_+}{\det_-}\,.
\end{equation}
Since this quantity is normalisation independent, we are free to renormalise any of the constituent states in any way we like. For example, although we have not introduced any complex structure, in such a scenario one could normalise $\langle \Psi|$ and $|W \rangle$ to be hermitian conjugates of $|\Psi\rangle$ and $\langle W|$, as well as impose $\langle W|W\rangle=1$. Then, \eqref{normalisationindep} would read
\begin{equation}
    \frac{|\langle W|\Psi\rangle|^2}{||\Psi||} = \frac{1}{3^{L}Q_\theta(\tfrac{i}{4})Q_\theta(-\tfrac{i}{4})} \frac{\det_+}{\det_-}
\end{equation}
where $||\Psi||$ denotes the norm of $|\Psi\rangle$ in the imposed complex structure.

\section{Principal Operator Insertions}\label{sec:fsov_insertions}

In the previous section we saw how the overlap $\langle W|\Psi\rangle$ could be computed in determinant form. We now turn to the computation of this overlap with insertions of operators. Namely, we will compute overlaps 
\begin{equation}
    \langle W|\sfP_{s,r}^\gamma(v)|\Psi\rangle
\end{equation}
where $\sfP_{s,r}^\gamma(v)$ is a two-parameter family of operators depending on complex numbers $v$ and $\gamma$ in a polynomial way as well as generalisations thereof. More precisely, $\sP_{a,r}^{\gamma}$ are essentially the \emph{Principal Operators} introduced in \cite{Gromov:2022waj}. To define standard Principal Operators of \cite{Gromov:2022waj} we use the fact that the integrals of motion are linear in the characters of the twist matrix (at least in the companion twist frame, see Appendix \ref{app:MicroscopicDef}):
\begin{equation}
    \hat{I}_{s,n} = \sum_{r=\pm 1,3}\chi_{r} \,\hat{I}^{(r)}_{s,n}\,.
\end{equation}
The Principal Operator is the generating series of the operators $\hat{I}^{(r)}_{s,n}$:
\begin{equation}
    \sP_{s,r}(u) = \delta_{r,s}\left(u+s\frac{\ii}{4}\right)^{2L}+\sum_{n=0}^{2L-1} \left(u+s\frac{\ii}{4}\right)^{n} \hat{I}^{(r)}_{s,n}\,.
\end{equation}
The deformed $\sP_{s,r}^\gamma$ are instead defined as 
\begin{equation}\label{Popgamma}
    \sP_{s,r}^\gamma(u)=\delta_{r,s}\left(u+s\frac{\ii}{4}\right)^{2L}+\sum_{\beta=1}^{L} \left(u+s\frac{\ii}{4}\right)^{2\beta-2}I^{(r)\gamma}_{s,2\beta-2}
\end{equation}
where $\gamma$ is a generic complex number and we define
\begin{equation}
    I^{(r)\gamma}_{s,2\beta} =I^{(r)}_{s,2\beta}+ \left(\gamma+s\frac{i}{4}\right)I^{(r)}_{s,2\beta+1}\,.
\end{equation}
When $\gamma=u$, $\sP^\gamma_{s,r}$ reduces to $\sP_{s,r}$, so the former is indeed a deformation of the latter. We will clarify our reasons for introducing $\sP^\gamma_{s,r}$ below.

\paragraph{Notation for insertions.} To display the result of this section in a compact manner it is useful to use the shorthand notation introduced in \cite{Gromov:2022waj} which we now recall. We will write
\begin{equation}
    [o_{s,\beta}] = \det_{(a,\alpha),(s,\beta)} \bl \bQ_{1}(u) o_{s,\beta} \bQ_{a}(-u) \br_{\alpha}\,,
\end{equation}
with $o_{s,\beta}$ an operator. If we should need to specify the entries more explicitly we will give $o_{s,\beta}$ as a list
\begin{equation}
    [o_{s,\beta}] = [o_{1,1},o_{1,2},o_{1,L},\dots,o_{-1,1},\dots,o_{-1,L}]\;.
\end{equation}
An example of this notation is the overlap $\braket{W}{\Psi}$  \eq{WPsioverlap} which we can now write in two equivalent ways as
\begin{equation}
    {\rm det}_+  = [D^{\frac{t}{2}}w^{2j-1}D^{\frac{t}{2}}] = [\{D^{\frac{1}{2}}w^{2\beta-1}D^{\frac{1}{2}}\}_{j=1}^{L},\{D^{-\frac{1}{2}}w^{2j-1}D^{-\frac{1}{2}}\}_{j=1}^{L}]\,.
\end{equation}
We have also replaced $u$ with $w$, $\beta$ with $j$ and $s$ with $t$ in the operators to make it easier to avoid confusion. Finally, it will be very useful to introduce the following replacement notation
\begin{equation}
    [(s,\beta)\rightarrow o] = [D^{\frac{1}{2}}w D^{\frac{1}{2}},\dots,D^{\frac{s}{2}}w^{2\beta-3} D^{\frac{s}{2}},o,D^{\frac{s}{2}}w^{2\beta+1} D^{\frac{s}{2}},\dots]\;.
\end{equation}  
Or in other words, $[(s,\beta)\rightarrow o]$ is the determinant of the matrix $D^{\frac{t}{2}}w^{2j-1}D^{\frac{t}{2}}$ with $D^{\frac{s}{2}}w^{2\beta-1}D^{\frac{s}{2}}$ replaced with $o$. Multiply insertions will be given as a list of replacement rules.

\subsection{Single Insertions}
The starting point is once again a trivial identity following from the Baxter equation. We generalise the identity \eqref{eqn:trivdiff1} by including additional prefactors $f(u),g(u)$ to find
\begin{equation}\label{eq:BaxterInsertionsFG}
    \bl Q^{A}_{1}\left(\overrightarrow{\mathcal{O}}_{A}\,f(u)- g(u)\overrightarrow{\mathcal{O}}_{\bar{A}}\right)Q^{a}_{\bar{A}} \br_{\alpha} = 0\,.
\end{equation}
When $f$ and $g$ are not simply the identity function there will no longer be a cancellation between terms that are independent of the states, such as $Q^-_{\theta}D^{3}Q^+_{\btheta}$. To make contact with the determinants in \eqref{eqn:spdet2} we should pick $f,g$ in such a way as to only get terms of the form $(u\pm \frac{\ii}{4})^{\text{odd}}$ when expanding out the equation in terms of integrals of motion. This fixes the functions to be
\begin{equation}
    f = \frac{u}{2}+\frac{\gamma}{2}\;, \quad g=\frac{u}{2}-\frac{\gamma}{2}\;,
\end{equation}
where $\gamma$ is an arbitrary constant. Rewriting \eqref{eq:BaxterInsertionsFG} then yields
\begin{equation}\label{eq:InsertionsExpansion}
\begin{split}
   & \displaystyle \sum_{\beta,s}s\;\bl \bQ_1(u) D^{\frac{s}{2}} u^{2\beta-1} D^{\frac{s}{2}}\bQ_a(-u)  \br_{\alpha}\; I^{\gamma}_{s,2\beta-2} = \sum_{s=\pm 1,3} \chi_{r}\bl \bQ_{1}(u)\mathcal{O}_{(r)}\bQ_{a}(-u)\br_{\alpha}\,,
\end{split}
\end{equation}
where we have collected together odd and even charges according to
\begin{equation}\label{eq:Igamma}
    I^{\gamma}_{s,2\beta} =I_{s,2\beta}+ (\gamma+s\tfrac{i}{4})I_{s,2\beta+1}\;.
\end{equation}
Finally, the operators on the r.h.s of \eqref{eq:InsertionsExpansion} are defined as
\begin{subequations}
\begin{align}
    &\mathcal{O}_{(3)} = \,Q_{\theta}^-D^{\frac{3}{2}}uD^{\frac{3}{2}}Q_{\btheta}^+\,,
    &
    &\mathcal{O}_{(-3)} = -Q_{\theta}^+D^{-\frac{3}{2}}uD^{-\frac{3}{2}}Q_{\btheta}^-\,,
    \\
    &\mathcal{O}_{(1)} = -D^{\frac{1}{2}}u^{2L+1}D^{\frac{1}{2}}\,,
    &
    &\mathcal{O}_{(-1)} = D^{-\frac{1}{2}}u^{2L+1}D^{-\frac{1}{2}}\,.
\end{align}
\end{subequations}
Due to our choice of $f$ and $g$ the coefficients on the l.h.s. of \eqref{eq:InsertionsExpansion} are exactly the matrix entries in the determinant formula for the overlap. \eqref{eq:InsertionsExpansion} is simply a standard matrix equation $({\rm det}_+)_{A}{}^{B}I^{\gamma}_{B} = (\mathcal{O}_{(r)})_{A}$ with multi-indices $A=\{\alpha,a\},B=\{s,\beta\}$ and $({\rm det}_+)_{\alpha,a}{}^{s,\beta} = \bl \bQ_{1} D^{\frac{s}{2}}w^{2\beta-1}D^{\frac{s}{2}} \bQ_{a}(-u)\br_{\alpha}$. We can then contract with the minor $\epsilon^{A_1\dots,A_kA}\epsilon_{B_1\dots B_kB} d_{A_1}{}^{B_1} \dots d_{A_k}{}^{B_k}$ to isolate $I^{\gamma}_{s,2\beta-2}$. We find
\begin{equation}\label{beforeproj}
    \mel{W}{\hat{I}^{\gamma}_{s,2\beta-2}}{\Psi} = s\sum_{r=\pm 1,3} \chi_{r} [(s,\beta)\rightarrow \mathcal{O}_{(r)}]\,,
\end{equation}
where we have used that $[D^{\frac{t}{2}}w^{2j-1}D^{\frac{t}{2}}]I^{\gamma}_{s,2\beta-2} = \langle{W}|{\Psi}\rangle  I^{\gamma}_{s,2\beta-2}= \mel{W}{\hat{I}^{\gamma}_{s,2\beta-2}}{\Psi}$. The now crucial observation, originally made in \cite{Gromov:2022waj} in the case of scalar products, is that after expanding $\hat{I}^{\gamma}_{s,2\beta-2}$ into $\hat{I}^{(r),\gamma}_{s,2\beta-2}$ we are allowed to match the coefficients of the characters $\chi_{r}$. This is a non-trivial statement since there is still twist dependence in the eigenvectors $\ket{\Psi}$ and in the Q-functions featuring in the determinant. The proof is analogous to the one that can be found in \cite{Gromov:2022waj}, for completeness, we recall it in Appendix~\ref{app:proofs_multiple_insertions}. The final result is
\begin{equation}\label{eq:expOfPrincipal}
    \mel{W}{\hat{I}^{(r),\gamma}_{s,2\beta-2}}{\Psi} =\mel{W}{\hat{I}^{(r)}_{s,2\beta-2}}{\Psi}= s\,[(s,\beta)\rightarrow \mathcal{O}_{(r)}]\,.
\end{equation}
where we have used that the r.h.s. does not depend on $\gamma$, which is also consistent with the fact that all odd charges are annihilated by $\bra{W}$.

\paragraph{Principal Operators.} Let us now collect $\hat{I}^{(r)}_{s,2\beta-1}$ into a generating operators, namely the principal operators $\sP^{\gamma}_{s,r}$ defined in \eqref{Popgamma}. To write down the expectations value of $\sP^{\gamma}_{s,r}$ it is very convenient to introduce the following notation
\begin{equation}
    \begin{split}
    &[L_0;\mathbf{u}_0\Big|L_1;\mathbf{u}_1|L_2;\mathbf{u_0} \Big| L_{3};\mathbf{u}_0] =
\[\left\{Q_{\theta}^-D^{\frac{3}{2}}\frac{\Delta^+_{\mathbf{u}_0 \cup w}}{\Delta^+_{\mathbf{u_0}}}w^{2\beta-2}D^{\frac{3}{2}}Q_{\btheta}^+\right\}_{\beta=1}^{L_0-1},\right. \\
    &\left.\quad \left\{D^{\frac{1}{2}}\frac{\Delta^+_{\mathbf{u}_1 \cup w}}{\Delta^+_{\mathbf{u}_1}}w^{2\beta-2}D^{\frac{1}{2}}\right\}_{\beta=1}^{L_1-1},\left\{D^{-\frac{1}{2}}\frac{\Delta^+_{\mathbf{u}_2 \cup w}}{\Delta^+_{\mathbf{u}_2}}w^{2\beta-2}D^{-\frac{1}{2}}\right\}_{\beta=1}^{L_2-1},\left\{Q_{\theta}^+D^{-\frac{3}{2}}\frac{\Delta^+_{\mathbf{u}_3 \cup w}}{\Delta^+_{\mathbf{u}_3}}D^{-\frac{3}{2}}Q_{\btheta}^-\right\}_{\beta=1}^{L_3-1}\]\,.
    \end{split}
\end{equation}
where $\Delta^+(\{A_i\}_{i}) = \prod_i A_i \prod_{i<j} (A^2_i-A^2_j)$.
For example we can calculate the expectation value of $\sP_{+,1}$:
\begin{equation}
\begin{split}
    \mel{W}{\sP^\gamma_{+,1}}{\Psi} &= \left(u+\frac{\ii}{4}\right)^{2L}\chi_{+}\braket{W}{\Psi}-\sum_{\beta=1}^{L} (u+\frac{\ii}{4})^{2\beta-2}\,[(+,\beta)\rightarrow D^{\frac{1}{2}}w^{2L+1}D^{\frac{1}{2}}]\\
    &=[\{D^{\frac{1}{2}}\left((u+\frac{\ii}{4})^2-w^2\right)w^{2\beta-1}D^{\frac{1}{2}}\}_{\beta=1}^{L},\{D^{-\frac{1}{2}}w^{2\beta-1}D^{-\frac{1}{2}}\}_{\beta=1}^{L}] \\
    &=[0;\big|L; u+\frac{\ii}{4}\big| L;0 \big| 0; ]\,.
\end{split}
\end{equation}
Where we can set $\gamma$ to $0$ as the r.h.s. does not depend on it.
For the remaining principal operators we have, using notation $\expval{f}_{W} = \mel{W}{f}{\Psi}$,
\begin{subequations}\la{Pinu}
\begin{alignat}{4}
&\expval{\sP_{+,3}}_{W} &&= &&(-1)^{L-1}&&[1;\big|\parbox{\widthLOne}{$L-1$};\parbox{\widthLTwo}{$u+\tfrac{\ii}{4}$}\big|\parbox{\widthLOne}{$L$};\parbox{\widthLTwo}{\,}\big|0;]\,, \\
    &\expval{\sP_{+,1}}_{W} &&= && &&[0;\big|\parbox{\widthLOne}{$L$};\parbox{\widthLTwo}{$u+\tfrac{\ii}{4}$}\big|\parbox{\widthLOne}{$L$};\parbox{\widthLTwo}{\,}\big|0;] \,,\\
    &\expval{\sP_{+,-1}}_{W} &&= &&(-1)^{L}&&[0;\big|L-1;\parbox{\widthLTwo}{$u+\tfrac{\ii}{4}$}\big|\parbox{\widthLOne}{$L+1$};\parbox{\widthLTwo}{\,}\big|0;]\,,\\
    &\expval{\sP_{+,-3}}_{W} &&= &&(-1)^{L+1}&&[0;\big|L-1;\parbox{\widthLTwo}{$u+\tfrac{\ii}{4}$}\big|\parbox{\widthLOne}{$L$};\parbox{\widthLTwo}{\,}\big|1;]\,, \\
    &\expval{\sP_{-,+3}}_{W} &&= &&&&[1;\big|\parbox{\widthLOne}{$L$};\parbox{\widthLTwo}{\,}\big|L-1;u-\tfrac{\ii}{4}\big|0;]\,, \\
    &\expval{\sP_{-,+1}}_{W} &&= &&(-1)^{L-1}&&[0;\big|\parbox{\widthLOne}{$L+1$};\parbox{\widthLTwo}{\,}\big|\parbox{\widthLOne}{$L-1$};u-\tfrac{\ii}{4}\big|0;]\,, \\
    &\expval{\sP_{-,-1}}_{W} &&= && &&[0;\big|\parbox{\widthLOne}{$L$};\parbox{\widthLTwo}{\,}\big|\parbox{\widthLOne}{$L$};u-\tfrac{\ii}{4}\big|0;]\,, \\
    &\expval{\sP_{-,-3}}_{W} &&= && &&[0;\big|\parbox{\widthLOne}{$L$};\parbox{\widthLTwo}{\,}\big|\parbox{\widthLOne}{$L-1$};u-\tfrac{\ii}{4}\big|1;] \,.
\end{alignat}
\end{subequations}

\subsection{Multiple Insertions}
The above considerations can be generalised to allow for multiple insertions. To do so we once again start from \eqref{eq:InsertionsExpansion} and contract it with an appropriate combination of $({\rm det}_+)_A{}^{B}$ and insertions of other operators. For example, in the case of $2$ insertions we use $\epsilon^{A_1\dots A_k A}\epsilon_{B_1 \dots B' B} ({\rm det}_+)_{A_1}{}^{B_1} \dots (\mathcal{O}_{(r')})_{A_k}$ to find
\begin{equation}
\begin{split}
    s[(s',\beta')\rightarrow \mathcal{O}_{(r')}] I^{\gamma}_{s,2\beta-2} - &s'[(\beta,s)\rightarrow \mathcal{O}_{(r')}] I^{\gamma}_{s',2\beta'-2} \\
    &= [(s',\beta')\rightarrow \mathcal{O}_{(r')},(s,\beta)\rightarrow \sum_{r}\chi_{r} \mathcal{O}_{(r)}]\,,
\end{split}
\end{equation} 
which we can rewrite in terms of operators as
\begin{equation}\label{eq:W2IPsi}
    \mel{W}{\hat{I}^{(r')\gamma'}_{s',2\beta'-2}\hat{I}^{\gamma}_{s,2\beta-2}-\hat{I}^{(r')\gamma'}_{s,2\beta-2}\hat{I}^{\gamma}_{s',2\beta'-2}}{\Psi} = s s'[(s',\beta')\rightarrow \mathcal{O}_{r'},(s,\beta)\rightarrow\sum_{r} \chi_{r}\mathcal{O}_{r}]\,.
\end{equation}
As opposed to the single-insertion formula we can in this case not drop the $\gamma$-superscript, it is crucial to amend the original odd charges according to \eqref{eq:Igamma} in order for \eqref{eq:W2IPsi} to be valid. This formula is furthermore amendable to the character projection trick described in the previous subsection using which we can conclude
\begin{equation}\la{doubleins}
    \mel{W}{\hat{I}_{s,2\beta-2}^{(r),\gamma}\hat{I}^{(r'),\gamma'}_{s',2\beta'-2}-\hat{I}_{s',2\beta'-2}^{(r),\gamma}\hat{I}^{(r'),\gamma'}_{s,2\beta-2}}{\Psi} = s s'[(s,\beta)\rightarrow \mathcal{O}_{r},(s',\beta') \rightarrow \mathcal{O}_{r'}]\,.
\end{equation}
We can see that the r.h.s is independent of both $\gamma$ and $\gamma'$, meaning that the terms dependent on either $\gamma$ or $\gamma'$ would have vanishing r.h.s. in \eq{doubleins}.

The same argumentation remains valid for multiple insertions, the final result is obtained as
\begin{equation}\la{components}
    \mel{W}{\hat{I}^{(r_1),\gamma_1}_{[s_1,2\beta_1-2}\hat{I}^{(r_2),\gamma_2}_{s_2,2\beta_2-2}\dots \hat{I}^{(r_m),\gamma_m}_{s_m,2\beta_m-2]}}{\Psi} =\frac{1}{m!} \prod_{i=1}^{m} (s_i) \times [\{(s_i,\beta_i)\rightarrow \mathcal{O}_{(r_i)}\}_{i=1}^{m}]\,,
\end{equation}
where $[\dots]$ denotes antisymmetrisations \textit{e.g.} $I_{[a} J_{b]} = \frac{1}{2}(I_{a}J_b-I_b J_a)$.

One can convert expressions \eq{components} into generating functions like in \eq{Pinu} for multiple insertions as well. For examples see~\cite{Gromov:2022waj}.

\section{Generating All Integrable Boundary States}\label{genboundaries}
\label{sec:gen_integrable_states}
In the preceding sections we managed to compute the overlap of the integrable boundary state $\langle W|$ with integral of motion eigenstates $|\Psi\rangle$, and also managed to compute the overlap with insertions of principal operators in determinant form. These insertions, while simple, act on $\langle W|$ leading it outside the class of integrable boundary states. Hence we can ask, can we find any more integrable boundary states for which the overlaps are equally simple in our formalism? 

There is one obvious way to construct more integrable boundary states. Given any integrable boundary state $\langle B|$, we can always generate another $\langle B'|$ by acting on the first $\langle B|$ with some integral of motion, a transfer matrix, or even Baxter Q-operator -- anything which commutes with all integrals of motion and hence preserves the reflection property \eqref{Wdeftau}. This idea is not new and has been used very fruitfully in \cite{Piroli:2017sei,pozsgay2019integrable} to construct huge families of boundary states. 

We will now argue that for our twisted alternating spin chain, a \textit{complete} basis of all integrable boundary states can be generated in this way, by starting with $\langle W|$ and acting with transfer matrices $\hat{\tau}_+(u)$ evaluated at special rapidities $u$. 

\subsection{Counting of Integrable Boundary States}

As a first step we will count the number of possible integrable boundary states. Clearly, integrable boundary states form a vector space and so we count the dimension of this space. 

Any boundary state $\langle B|$ only has non-zero overlap with integral of motion eigenstates $|\Psi\rangle$ whose Q-functions satisfy the reflection rule $Q_a(-u) =(-1)^L Q^a(u)$, $a=1,2,3$. Since eigenstates form a basis and are in a one-to-one correspondence with Q-system solutions, we can bound the number of integrable boundary states by counting the number of solutions of the Q-system obeying the selection rule in some convenient limit. 

\paragraph{Singular twist limit.} It is convenient to take the singular twist limit $\lambda_1\gg \lambda_2 \gg \lambda_3$. In this limit the Q-system simplifies massively, producing 
\begin{equation}
    q_1^- q_2^+ =(-1)^L Q_\theta(u) q_3(-u),\quad q_1^- q_3^+ =(-1)^L Q_\theta(u) q_2(-u),\quad q_2^- q_3^+ =(-1)^L Q_\theta(u) q_1(-u)
\end{equation}
where we remind that $q_a$ denotes the polynomial part of the Q-function $Q_a = \lambda_a^{i u} q_a$. Counting the number of solutions in this regime is easy due to the factorised nature of the problem. For example, any root of $q_1^-$ must be a root of either $Q_\theta(u)$ or $q_2^+$, and so on. As a result of the factorisation, it is enough to count solutions for $L=1$. If there are $\#$ solutions, then for general $L$ there will be $\#^L$ solutions, assuming inhomogeneities $\theta_\alpha$ are not separated by an integer multiple of $i$. 

For $L=1$ the counting is trivial -- there are $3$ solutions. Hence, there are $3^L$ solutions for general $L$. 

We would then like to conclude that the space of integrable boundary states has dimension $3^L$. We should however be careful. The spectrum of the Q-system is non-degenerate in the singular twist limit \footnote{In the singular twist limit the Q-system describes the Gelfand-Tsetlin subalgebra, which has non-degenerate spectrum \cite{molev2007yangians}.}, so we do not run the risk of there being two distinct finite-twist solutions degenerating into the same singular twist solutions. Hence, there are at most $3^L$ solutions at finite twist. What we can not rule out is that the reflection invariance $Q_a(-u) = (-1)^L Q^a(u)$ is broken for a singular twist solution as we turn on corrections in twist. 

To conclude, all we can state is that the dimension of the space of integrable boundary states has dimension at most $3^L$. In the next subsection we will prove that it is precisely $3^L$, by explicitly constructing $3^L$ linearly independent boundary states.

\subsection{Basis of Integrable Boundary States}

We now aim to construct a family of integrable boundary states in the following way 
\begin{equation}
    \langle W| \displaystyle \prod_{i=1}\hat{\tau}_+(u_i)
\end{equation}
where $u_i$ are some rapidities. One thing we immediately notice is the similarity between our proposal and the construction of the separated variable basis \cite{Maillet:2018bim,Ryan:2018fyo,Maillet:2018czd}, which is built from an appropriate ``vacuum" $\langle 0|$ and repeatedly acting with transfer matrices. Since the dimension of the space is at most $3^L$, which is the dimension of an $\mathfrak{su}(3)$ spin chain of length $L$, we can try to repeat the prescription for the SoV basis when all sites are in the fundamental representation. Hence, we will label a state $\langle B_n|$ by a tuple $(n_1,\dots,n_L)$ with the definition
\begin{equation}\label{BdystateBn}
    \langle B_n| := \langle W| \displaystyle \prod_{\alpha=1}^L \left(\frac{\hat{\tau}_+(\theta_\alpha+i/2)}{Q_\theta^{[2]}(\theta_\alpha)Q_{\bar\theta}^{[3]}(\theta_\alpha)}\right)^{n_\alpha}.
\end{equation}
with $n_\alpha=0,1,2$. 

Clearly, all of these $3^L$ states are integrable boundary states for any choice of $n_1,\dots,n_L$. Remarkably, they are also all non-zero and linear independent! We prove these properties in Appendix \ref{app:basis_integrable_states}. Hence, we have proven that the dimension of the space of integrable boundary states is precisely $3^L$ and given an explicit construction for a basis.

We have chosen the normalistion factor in \eqref{BdystateBn} to simplify the overlap with transfer matrix eigenstates. Indeed, by definition we have
\begin{equation}
    \langle B_n|\Psi\rangle = \langle W|\Psi\rangle \displaystyle \prod_{\alpha=1}^L \left(\frac{\tau_+(\theta_\alpha+i/2)}{Q_\theta^{[2]}(\theta_\alpha)Q_{\bar\theta}^{[3]}(\theta_\alpha)}\right)^{n_\alpha}\,.
\end{equation}
Using the expression \eqref{eq:TauFromQsu3} for $\tau_+(u)$ in terms of Q-functions we then immediately obtain
\begin{equation}
    \langle B_n|\Psi\rangle = \langle W|\Psi\rangle \displaystyle \prod_{\alpha=1}^L \left(\frac{Q_1^-(\theta_\alpha)}{Q_1^+(\theta_\alpha)}\right)^{n_\alpha}\,.
\end{equation}

Similarly, a basis of right boundary states $|B_n\rangle$ can be constructed in an identical way, for example, we can take 
\begin{equation}
    |B_n\rangle = \displaystyle \prod_{\alpha=1}^L \left(\frac{\hat{\tau}_-(\theta_\alpha-i/2)}{\chi_{-3}Q_\theta^{[-2]}(\theta_\alpha)Q_{\bar\theta}^{[-3]}(\theta_\alpha)}\right)^{n_\alpha} |W\rangle\;,
\end{equation}
with an extra $\chi_{-3}$ factor appearing due to the fact that we have moved twist factors around by replacing $\hat{\tau}_+ \rightarrow \hat{\tau}_-$. The overlap with $\langle \Psi|$ is then computed to be 
\begin{equation}
    \langle \Psi|B_n\rangle = \langle \Psi|W\rangle \displaystyle \prod_{\alpha=1}^L \left(\frac{Q_1^+(\theta_\alpha)}{Q_1^-(\theta_\alpha)}\right)^{n_\alpha}\,.
\end{equation}

The overlap $\langle B_n|B_n\rangle$ is remarkably simple to compute. Indeed, one just needs to use the fact that the product $\hat{\tau}_+(u+i/2)\hat{\tau}_-(u-i/2)$ becomes proportional to the identity operator at special values of the spectral parameter $u$. For us, we have \cite{molev2007yangians}
\begin{equation}
    \hat{\tau}_+(\theta_\alpha+i/2)\hat{\tau}_-(\theta_\alpha-i/2) = \lambda_1 \lambda_2 \lambda_3\,Q_\theta^{[2]}(\theta_\alpha)Q_\theta^{[-2]}(\theta_\alpha)Q_{\bar\theta}^{[3]}(\theta_\alpha)Q_{\bar\theta}^{[-3]}(\theta_\alpha)
\end{equation}
and as a result, we have
\begin{equation}
    \langle B_n|B_n\rangle = \langle W|W\rangle
\end{equation}
for all $n$, with $\langle W|W\rangle$ computed in \eqref{Wnorm}. In a similar way, we can see that
\begin{equation}
    \langle B_n|\Psi\rangle \langle \Psi|B_n\rangle = \langle W|\Psi\rangle \langle \Psi|W\rangle
\end{equation}
and as a result the normalisation independent quantity \eqref{normalisationindep} is invariant under the replacement $W\rightarrow B_n$. 

\section{Outlook}\la{sec:outlook}

In this paper we showed how the problem of computing overlaps with integrable boundary states is naturally encoded in the Functional Separation of Variables (FSoV) formalism.

There are numerous possible directions for generalisations:

\begin{itemize}
    \item Different representations in the physical space, especially non-compact ones \cite{Jiang:2020sdw}. The SoV wave functions are known for all finite-dimensional representations in the physical space \cite{Ryan:2018fyo,Ryan:2020rfk} and so it should be simple to generalise the FSoV construction. FSoV is already known in the non-compact set-up \cite{Gromov:2020fwh}. Representation-independent overlap formulas were recently obtained for any highest weight representation using the Bethe ansatz approach \cite{Gombor:2023bez}, and it would be interesting if such formulas could be obtained using FSoV. 

    \item How specific boundaries are expressed in terms of the basis states $\langle B_n|$, in particular how to relate them to solutions of the KT-relations \cite{Frassek:2017fba,Gombor:2021uxz,Gombor:2023bez,Gombor:2021hmj} or crosscap states \cite{Gombor:2022deb,Caetano:2021dbh,Ekman:2022kpx}.

    \item Removing the boundary twist. We focused on spin chains with twisted boundary conditions to ensure regularity of the SoV construction. For untwisted spin chains there is a far larger class of integrable boundary states, for example Matrix Product States which are generically non-integrable in the presence of twist \cite{Widen:2018nnu}. Owing to their physical relevance to one-point functions in defect $\lN=4$ SYM it would be very interesting to determine if overlaps with them can also be obtained from the $\langle W|\Psi\rangle$ overlap by action of some transfer matrices of Q-operators. Another possible avenue is to relate them to $\langle W|$ by the action of principal operators. As we have described in the text these states are not integrable, but their overlaps with $|\Psi\rangle$ are still extremely simple.

    \item XXZ or $q$-deformed spin chains, where it should be possible to compare with \cite{pozsgay2014overlaps,pozsgay2018overlaps}, and supersymmetric systems, where we should be able to compare with \cite{Kristjansen:2020vbe,Kristjansen:2021xno}.

    \item In the language of \cite{Gombor:2021hmj,Gombor:2023bez} the boundary states we consider here are referred to as ``untwisted" as the reflection property relates the same transfer matrices $\hat{\tau}_\pm \rightarrow \hat{\tau}_\pm$. There is another type of reflection, called ``twisted" which swaps the transfer matrices $\hat{\tau}_\pm \rightarrow \hat{\tau}_\mp$. Wilson loops are one example of such a boundary state \cite{Cavaglia:2021mft} and FSoV was already used to conjecture the universal part of those overlaps in fishnet theory. With the methods developed here we should be able to prove it. 

    \item Thermodynamic limit. This direction is important for understanding how techniques such as FSoV can be applied to integrable quantum field theories. It is already known that the universal part of boundary overlaps is given in terms of Fredholm determinants \cite{Caetano:2020dyp}, so we expect our determinant expression for $\langle W|\Psi\rangle$ to take such a form in the thermodynamic limit. See \cite{Gromov:2016itr} for the related question of relating determinants of Q-functions to the Gaudin determinant.

    \item Correlators with two determinant operators and a single-trace local operator in planar $\lN=4$ SYM take the form of overlaps with integrable boundary states \cite{Jiang:2019zig,Jiang:2020sdw}. Can we combine our results with the higher loop SoV results of \cite{Bercini:2022jxo} to compute such overlaps and compare with \cite{Buhl-Mortensen:2017ind}? Alternatively can we do it in the fishnet theory where FSoV is known at all loops \cite{Cavaglia:2021mft} and make contact with \cite{Shahpo:2021xax}?
\end{itemize}

\paragraph{Acknowledgements} We are grateful to K. Zarembo for discussions and M. de Leeuw for comments on the draft. The work of S.E, N.G. and P.R was
supported by the European Research Council (ERC) under the European Union’s Horizon 2020 research and innovation program – 60 – (grant agreement No. 865075) EXACTC.

\appendix

\section{Microscopic Formulation for $\su(3)$ }\label{app:MicroscopicDef}
In this appendix, we provide the standard construction of the alternating $\mathfrak{su}(3)$ spin chain considered in the main text using the RTT-formalism.
\subsection{Lax Operators}

To begin with, we will construct representations of the Lie algebra generators for both $\mathbf{3}$ and $\overline{\mathbf{3}}$.

\paragraph{Lie algebra generators and representation.}

The Lie algebra generators $\bbE_{ij}$ of $\gl(3)$ satisfy the commutation relations
\begin{equation}\label{eqn:Liealg}
    [\bbE_{ij},\bbE_{kl}]=\delta_{jk}\, \bbE_{il} - \delta_{li}\, \bbE_{kj}\,.
\end{equation}

First we consider the defining representation ${\bf 3}$ of $\frak{gl}(3)$. We introduce the standard matrices $\mathsf{e}_{ij}$ which are $3\times 3$ matrices with $1$ at position $i,j$ and all other entries set to $0$. Using these matrices we set $\mathbb{E}_{ij} = \sfe_{ij} - \frac{1}{2}\delta_{ij}$. We will denote the standard basis of $\mathbb{C}^3$ by $\sfe_i$. These vectors are defined in the same way as $\sfe_{ij}$ and satisfy $\sfe_{ij}\sfe_k = \delta_{jk} \sfe_i$. We write the generators of the dual representation $\overline{\mathbf{3}}$ as $\overline{\mathbb{E}}_{ij} = -\mathsf{e}_{4-j,4-i}+\delta_{ij}$ and define the dual basis vectors $\overline{\sfe}_i = \sfe_{4-i}$. We have introduced the overall shifts with $\delta_{ij}$ to simplify the expressions in the next sections.

\paragraph{Lax operators and monodromy matrix.}

We will now proceed to construct the model, whose integrability is guaranteed by a Lax representation. We construct two Lax operators, $\lL$ and $\bar{\lL}$ given by
\begin{equation}
    \lL_{jk}(u) = u\, \delta_{jk} + i\, \bbE_{kj},\quad \bar{\lL}_{jk}(u) = u\, \delta_{jk} + i\, \bar{\bbE}_{kj}\,.
\end{equation}
From here, we will build the monodromy matrix for a spin chain of length $2L$, with $L$ sites in the $\bf 3$ and $L$ sites in the $\bf\bar{3}$
\begin{equation}\label{eqn:monod}
    T_{ij}(u) = \displaystyle \sum_{k} \lL^{(1)}_{i k_1}(u-\theta_1)\bar{\lL}^{(2)}_{k_1 k_2}(u-\bar{\theta}_1)\dots \lL^{(2L-1)}_{k_{2L-2} k_{2L-1}}(u-\theta_L)\bar{\lL}^{(2L)}_{k_{2L-1} j}(u-\bar{\theta}_L)
\end{equation}
where all indices $k_1,\dots, k_{2L-1}\in\{1,2,3\}$ are summed over, $\theta_\alpha$ are inhomogenieties and we use the notation $\bar{\theta}_\alpha=-\theta_\alpha$. The inhomogeneities are assumed to be generic, in particular they are not related by integer multiples of $i$. We further introduce the twisted monodromy matrix
\begin{equation}
    \bfT_{ij}(u) = \sum_{k}T_{ik}(u)\Lambda_{kj}\;,
\end{equation}
where $\Lambda$ is the boundary twist matrix. For the most part, its precise form will not be hugely important. For us, the most important requirement is that it has distinct non-zero eigenvalues $\lambda_1$, $\lambda_2$, $\lambda_3$. We will also make use of symmetric combinations (characters) of these eigenvalues, which we denote as 
\begin{equation}\label{eqn:characters}
    \chi_{+3}:=1\;,\quad \chi_{+1} = \lambda_1 + \lambda_2 +\lambda_3\;,\quad \chi_{-1} = \lambda_1 \lambda_2 + \lambda_1 \lambda_3 +\lambda_2 \lambda_3\;,\quad \chi_{-3} = \lambda_1 \lambda_2 \lambda_3\;.
\end{equation}
For convenience we will sometimes just denote $\chi_{\pm 1}=\chi_\pm$. Note that because of the global $GL(3)$ symmetry different choices of $\Lambda$ with the same eigenvalues do not affect any physical observables -- all results for the spectrum of conserved quantities and the matrix elements of operators is invariant. Hence when performing calculations we can choose any $\Lambda$ we find convenient. We pick $\Lambda$ to be in the so-called companion twist form
\begin{equation}
    \Lambda = \begin{pmatrix}
        \chi_{+} & -\chi_{-} & \chi_{-3} \\
        1 & 0 & 0 \\
        0 & 1 & 0 
    \end{pmatrix}
\end{equation}
which is most convenient for SoV calculations, see for example \cite{Ryan:2018fyo,Ryan:2020rfk,Gromov:2022waj}. Note, that in the standard matrix basis $\Lambda$ is written
\begin{equation}\label{Lambdadefiningbasis}
    \Lambda= \chi_+ \sfe_{11}-\chi_-\sfe_{12}+\chi_{-3}\sfe_{13}+\sfe_{21}+\sfe_{32}\,. 
\end{equation}

\subsection{Conserved Charges and Transfer Matrices}

The most straightforward way to build conserved charges of the model is to take the trace of the monodromy matrix, obtaining the fundamental transfer matrix 
\begin{equation}
    \T_+(u) = \displaystyle \sum_i \bfT_{ii}(u)\;.
\end{equation}
The coefficients in the $u$-expansion of $\T_+(u)$ generate a commutative family of charges
\begin{equation}
    [\T_+(u),\T_+(v)]=0\;.
\end{equation}
However, this is not the only set of conserved charges we can construct.  We can construct more using the fusion procedure. To this end, we define anti-symmetrised combinations of monodromy matrix elements as 
\begin{equation}
    \bfT\left[^{i_1 i_2}_{j_1 j_2}\right](u) = \bfT_{i_1 j_1}\left(u+\frac{i}{2}\right)\bfT_{i_2 j_2}\left(u-\frac{i}{2}\right) - \bfT_{i_1 j_2}\left(u+\frac{i}{2}\right)\bfT_{i_2 j_1}\left(u-\frac{i}{2}\right)\;.
\end{equation}
This quantity is manifestly anti-symmetric in the $j_1,j_2$ indices. It is a non-trivial consequence of the RTT relations that it is also anti-symmetric in the upper indices. With these objects, we can build a new transfer matrix $\T_-(u)$ defined by 
\begin{equation}
    \T_-(u) = \bfT\left[^{12}_{12}\right](u) + \bfT\left[^{13}_{13}\right](u)+\bfT\left[^{23}_{23}\right](u)\,.
\end{equation}
The coefficients in the $u$-expansion of $\T_-(u)$ mutually commute with themselves and with those of $\T_+(u)$
\begin{equation}
    [\T_-(u),\T_-(v)] = [\T_-(u),\T_+(v)]=0
\end{equation}
guaranteeing that they form a mutually commutative family of conserved charges. 

\paragraph{Removing trivial factors}

By construction, $\T_+(u)$ is a polynomial of degree $2L$ and $\T_-(u)$ is a polynomial of degree $4L$. However, not all of the conserved charges in $\T_-(u)$ are independent and $\T_-(u)$ has a number of trivial zeroes. We can factor these trivial zeroes out by introducing the degree $2L$ operator $\hat{\tau}_-(u)$ defined by 
\begin{equation}
   \T_-(u) = \hat{\tau}_-\left(u+\frac{i}{2}\right) Q_\theta^{[-2]}Q_{\bar\theta}^{[3]}\;,
\end{equation}
and we recall from the main text that $Q_{\theta}= \prod(u-\theta_\alpha)$, $Q_{\btheta}=\prod(u-\theta_{\bar{\alpha}})$. To keep notation consistent, since $\T_+(u)$ does not have any trivial zeroes we will simply write $\T_+(u)=\hat{\tau}_+(u)$.

Finally, we note that the characters \eqref{eqn:characters} of the twist matrix are encoded in the large-$u$ asymptotics of the transfer matrices\footnote{The character $\chi_{-3}$ is encoded in the transfer matrix $\T_{-3}(u)$, corresponding to a totally anti-symmetrised combination of monodromy matrices (quantum determinant). We will not have any direct use for the explicit form of this object, so we do not write it down.}
\begin{equation}
    \hat{\tau}_\pm(u) = \chi_\pm\, u^{2L} + \dots
\end{equation}

\paragraph{Eigenstates} For the class of representations we consider the transfer matrices $\hat{\tau}_{\pm}(u)$ have non-degenerate spectrum. We will denote joint eigenstates of them by $|\Psi\rangle$. Their eigenvalues will be denotes by $\tau_+(u)$ and $\tau_-(u)$, i.e. 
\begin{equation}
    \hat{\tau}_+(u) |\Psi\rangle = \tau_+(u) |\Psi\rangle\;,\quad \hat{\tau}_-(u) |\Psi\rangle = \tau_-(u) |\Psi\rangle\;.
\end{equation}
When convenient we will single out the eigenstate
\begin{equation}\label{eq:OmegaDefAppendix}
    \ket{\Omega} \propto \bigotimes^{L} \left(\sfe_{1}+\frac{1}{\lambda_1}\sfe_2+\frac{1}{\lambda_1^{2}}\sfe_{3} \right)\otimes \left(\bar{\sfe}_{3}-\frac{\lambda_1+\lambda_2}{\lambda_1\lambda_2}\bar{\sfe}_{2}+\frac{1}{\lambda_1\lambda_2}\bar{\sfe}_1 \right)
\end{equation}
as a ground state of our construction. This is natural since the eigenvalues $\tau_{\pm}$ for this state are given by \eqref{eq:TauFromQsu3} with trivial $Q_{1}$ and $Q^{3}$, i.e $Q_{1} = \lambda^{\ii u},Q^{3} = \lambda^{-\ii u}_{3}$. The fact that this state appears complicated is just a consequence of the fact that we work with the companion twist -- if we were to rotate this back to the diagonal twist frame the state $|\Omega\rangle$ would simply be $\bigotimes^{2L}\sfe_1$.

\section{Separation of Variables Bases}\la{app:sov}

\subsection{$\langle \svx|$ Basis}

The SoV basis $\langle \svx|$ for the representations considered in this paper was constructed in \cite{Ryan:2018fyo,Ryan:2020rfk}, using a generalised version of the construction proposed in \cite{Maillet:2018bim} for the set-up where all sites are in the defining representation. The basis elements $\bra{\svx}$ are eigenstates of the operator $\mathbf{B}$ \cite{Gromov:2016itr}, defined in the companion twist frame as \cite{Ryan:2018fyo}
\begin{equation}
    \mathbf{B} = T_{11}(T_{11}T_{22}^{[-2]}-T_{12}T_{21}^{[-2]})+(T_{11}T_{23}^{[-2]}-T_{13}T_{21}^{[-2]})T_{21}\,.
\end{equation}
$\mathbf{B}$ has three key properties. First, it is manifestly independent of the twist eigenvalues when we are in the companion twist frame, and hence so are its eigenvectors. It commutes with itself at different values of the spectral parameter $u,u'$: $[\mathbf{B}(u),\mathbf{B}(u')]=0$. Finally, it has an explicitly computable spectrum
\begin{equation}
    \bra{\sx}\mathbf{B}(u) = Q_{\btheta}^{[2]}Q_{\theta}^{[-3]}\prod_{\alpha=1}^{L}\prod_{a=1}^{2}(u-\sx_{\alpha,a})(u-\bsx_{\alpha,a})\bra{\sx}\,,
\end{equation}
where $\sx_{\alpha,a},\bsx_{\alpha}$ are conveniently parameterised as
\begin{equation}
    \svx_{\alpha,a}=\theta_\alpha-\frac{i}{2}+i\, n_{\alpha,a}\;, 
      \quad
      \bar{\svx}_{\alpha,a}=\bar{\theta}_\alpha+i\, (\bar{n}_{\alpha,a}+1-a)\;,
\end{equation}
and $n_{\alpha,a}$ and $\bar{n}_{\alpha,a}$ are integers subject to the selection rules 
\begin{equation}
    1\geq n_{\alpha,1}\geq n_{\alpha,2}\geq 0\;,\quad 1\geq \bar{n}_{\alpha,1}\geq \bar{n}_{\alpha,2}\geq 0\;.
\end{equation}

The starting point for the basis construction is the vector $\langle 0|$, with all $n_{\alpha,a}=\bar{n}_{\alpha,a}=0$, referred to as the SoV vacuum, defined as, in the companion frame,
\begin{equation}
    \langle 0 | = \bigotimes^{2L}\left( \begin{array}{ccc}
        1 & 0 & 0 
    \end{array}\right)\,.
\end{equation}
Recall from Section~\ref{linkSoV} that we require the overlap between $\bra{\mathsf{x}}$ and $\ket{\Psi}$ to be given by
\begin{equation}\label{eqn:xwvApp}
    \langle \svx|\Psi\rangle =\displaystyle \prod_{\alpha=1}^L
    \oint_{\mathsf{x}_{\alpha,a}} \frac{du_1}{2\pi i}\frac{du_2}{2\pi i}\rho_\alpha(u_1)\rho_\alpha(u_2)    
\bQ_1(u_1)\bQ_1(u_2)\left[\bQ_1(\bar{\svx}_{\alpha,1})\bQ_3(\bar{\svx}_{\alpha,2})-
\bQ_1(\bar{\svx}_{\alpha,2})\bQ_3(\bar{\svx}_{\alpha,1})\right]
\end{equation}
This implies that we can now fully fix the normalisation of all eigenstates $\ket{\Psi}$.  This is possible because of the important property $\braket{0}{\Psi} \neq 0$ \cite{Ryan:2018fyo}.

The remaining SoV vectors $\langle \svx|$ can then, up to proportionality, be created using $\hat{\tau}_{\pm}$ as
\begin{equation}
    \langle \svx| \propto \langle 0| \displaystyle\prod_{\alpha=1}^L \hat{\tau}_-\left(\theta_\alpha-\frac{i}{2}\right)^{n_{\alpha,1}+n_{\alpha,2}}\,\hat{\tau}_+(\bar{\theta}_\alpha)^{\bar{n}_{\alpha,1}-\bar{n}_{\alpha,2}}\,\hat{\tau}_-(\bar{\theta}_\alpha-i)^{\bar{n}_{\alpha,2}}\,.
\end{equation}
The overall normalisation is fixed by requiring \eqref{eqn:xwvApp} using a convenient eigenstate of the transfer matrices, for example $\ket{\Omega}$ defined in \eqref{eq:OmegaDefAppendix}. The fact that these states reproduce the overlaps \eqref{eqn:xwvApp} follow as a consequence of the Baxter equations, see for example \cite{Ryan:2020rfk}
\subsection{$|\svy\rangle$ Basis}
The construction of the dual SoV basis $\ket{\svy}$ follows closely that of $\bra{\sx}$ \cite{Gromov:2019wmz,Gromov:2020fwh,Ryan:2021duf}. $\ket{\sy}$ are (right) eigenvectors of an operator $\mathbf{C}$ defined as\footnote{We define $\mathbf{C}$ with an additional shift as compared to previous studies. This additional shift simplifies its spectrum.}
\begin{equation}
    \mathbf{C}(u) = T_{11}^{-}(T_{11}^-T_{22}^{+}-T_{12}^-T_{21}^{+})+(T_{11}^-T_{23}^+-T_{13}^-T_{21}^+)T_{21}^{-}\,.
\end{equation}
with spectrum
\begin{equation}
    \mathbf{C}(u)\ket{\sy} = Q_{\btheta}^{[2]}Q_{\theta}^{[-3]}\prod_{\alpha=1}^{L}\prod_{a=1}^{2}(u-\sy_{\alpha,a})(u-\bar{\sy}_{\alpha,a}) \ket{\sy}\,.
\end{equation}
Here $\sy_{\alpha,a}$ and $\bar{\sy}_{\alpha,a}$ are given as
\begin{equation}
    {\svy}_{\alpha,a} = +{\theta}_\alpha + i\,({m}_{\alpha,a}+1-a)\;, \quad \bar{\svy}_{\alpha,a} = -\theta_\alpha + i\, \bm_{\alpha,a}-\frac{i}{2}\;,
\end{equation}
with $m_{\alpha,a},\bar{m}_{\alpha,a}$ integers subject to the selection rules
\begin{equation}
    1 \geq m_{\alpha,1} \geq m_{\alpha,2} \geq 0,\quad 1 \geq {\bm}_{\alpha,1} \geq {\bm}_{\alpha,2} \geq 0\,.
\end{equation}
We once again define a vacuum state
\begin{equation}
    \ket{0} = \bigotimes^{2L} \begin{pmatrix}
        1 \\ 0 \\ 0 
    \end{pmatrix}
\end{equation}
which in particular ensures the normalisation $\braket{0}{0}=1$. From here we can construct the remaining $|\svy\rangle$ states by 
\begin{equation}
    |\svy\rangle\, \propto\, \displaystyle \prod_{\alpha=1}^L \hat{\tau}_+(\theta_\alpha-i/2)^{m_{1,\alpha}-m_{2,\alpha}}\hat{\tau}_-(\theta_\alpha-i/2)^{m_{2,\alpha}}\hat{\tau}_+(\bar{\theta}_\alpha-i)^{\bar{m}_{1,\alpha}+\bar{m}_{2,\alpha}}|0\rangle
\end{equation}
where the normalisation of all left eigenstates $\bra{\Psi}$ and $\ket{\sy}$ are then finally to be fixed by requiring
\begin{equation}
    \langle \Psi| \svy\rangle =\displaystyle \prod_{\alpha=1}^L
    \oint_{\bar\svy_{\alpha,a}} \frac{du_1}{2\pi i}\frac{du_2}{2\pi i}\rho_{\bar\alpha}^+(u_1)\rho_{\bar\alpha}^+(u_2)    
\bQ^1(u_1)\bQ^1(u_2)\left[\bQ^1({\svy}_{\alpha,1})\bQ^3({\svy}_{\alpha,2})-
\bQ^1({\svy}_{\alpha,2})\bQ^3({\svy}_{\alpha,1})\right]\,.
\end{equation}

\subsection{Factorisable States}

Let's introduce some functions $G_1(u),G_3(u),G^1(u), G^3(u)$ (upper and lower indices here is just a labelling, there is no relation implied between them) and construct 
\begin{equation}
    {\bf G}_a(u) = \frac{G_a(u)}{Q_\theta^+ Q_\theta^-},\quad  {\bf G}^a(u) = \frac{G^a(u)}{Q_{\bar\theta}^+ Q_{\bar \theta}^-}\,.
\end{equation}
A factorisable state is any state $|\Theta\rangle$ whose wave function in the $\langle \svx|$ basis factorises as in \eqref{eqn:xwvApp}, i.e there exists some $G_1$ and $G_3$ such that
\begin{equation}
     \langle \svx|\Theta\rangle =\displaystyle \prod_{\alpha=1}^L
    \oint_{\mathsf{x}_{\alpha,a}} \frac{du_1}{2\pi i}\frac{du_2}{2\pi i}\rho_\alpha(u_1)\rho_\alpha(u_2)    
{\bf G}_1(u_1){\bf G}_1(u_2)\left[{\bf G}_1(\bar{\svx}_{\alpha,1}){\bf G}_3(\bar{\svx}_{\alpha,2})-
{\bf G}_1(\bar{\svx}_{\alpha,2}){\bf G}_3(\bar{\svx}_{\alpha,1})\right]\,.
\end{equation}
Similarly a state $\langle \Phi|$ is factorisable in the $|\svy\rangle$ basis if the exists some $G^1$ and $G^3$ such that
\begin{equation}
     \langle \Phi| \svy\rangle =\displaystyle \prod_{\alpha=1}^L
    \oint_{\bar\svy_{\alpha,a}} \frac{du_1}{2\pi i}\frac{du_2}{2\pi i}\rho_{\bar\alpha}^+(u_1)\rho_{\bar\alpha}^+(u_2)    
{\bf G}^1(u_1){\bf G}^1(u_2)\left[{\bf G}^1({\svy}_{\alpha,1}){\bf G}^3({\svy}_{\alpha,2})-
{\bf G}^1({\svy}_{\alpha,2}){\bf G}^3({\svy}_{\alpha,1})\right]\,.
\end{equation}
Obviously transfer matrix eigenstates are a clear example of factorisable states. However, they also describe ``off-shell" Bethe states -- states where $G_a$ are twisted-polynomials but whose roots do not satisfy the Bethe equations.

The overlap of two factorisable states can be easily computed using results following from FSoV. Indeed, we have
\begin{equation}
    \langle \Phi|\Theta\rangle = \displaystyle\sum_{\svx,\svy} \Phi(\svy)\lM_{\svy,\svx} \Theta(\svx)
\end{equation}
where the measure is fixed from FSoV. But we know the result takes the form of a determinant \eqref{eqn:spdet}, and so we can write (see for example \cite{Gromov:2020fwh})
\begin{equation}
    \langle \Phi|\Theta\rangle = \frac{1}{\cal N}\displaystyle\det_{(a,\alpha),(s,p)} \left(
    \begin{array}{c}
        \bl {\bf G}_1\; D^{\tfrac{s}2} \;u^{p-1}\; D^{\tfrac{s}2}\; {\bf G}^a\br_\alpha   \\
        \bl {\bf G}_a\; D^{\tfrac{s}2} \;u^{p-1}\; D^{\tfrac{s}2}\;{\bf G}^1\br_{\bar\alpha}  
    \end{array}
    \right)\;,
\end{equation}

In exactly the same way, we can compute the overlap of $\langle W|$ with any factorisable state
\begin{equation}
    \langle W|\Theta\rangle = \displaystyle\det_{(a,\alpha),(s,\beta)}         \bl {\bf G}_1(u)\; D^{\tfrac{s}2} \;u^{2\beta-1}\; D^{\tfrac{s}2}\; {\bf G}_a(-u)\br_{\alpha} \;.
\end{equation}

\section{Properties of $\langle W|$}\la{app:properties_w}

We will now collect some properties of the state $\langle W|$ in the $\mathfrak{su}(3)$ case. 

\subsection{Singlet Property}

\paragraph{Twist Independence.}

First, note that $\langle W|$ is completely independent of the twist matrix eigenvalues $\lambda_1, \lambda_2,\lambda_3$. Indeed, by explicit computation the overlaps $\langle W|\svx\rangle$ are independent of twist. Since the SoV basis $|\svx\rangle$ itself is independent of twist in the companion twist frame, $\langle W|$ is then also independent in this frame.  The transfer matrix $\hat{\tau}_+$ takes the explicit form 
\begin{equation}
   \hat{\tau}_+(u) = T_{12}(u) + T_{23}(u)+\chi_{+}\, T_{11}(u) - \chi_{-}\, T_{21}(u) + \chi_{-3}\, T_{31}(u)\,.
\end{equation}
Since $\langle W|$ in this frame is independent of $\lambda$s, and hence $\chi$s, the reflection invariance property of $\langle W|$
\begin{equation}
    \langle W|\hat{\tau}_+(u) = \langle W|\hat{\tau}_+\left(u-\frac{i}{2} \right)
\end{equation}
implies a far stronger set of conditions, namely 
\begin{equation}\label{eqn:newparity}
\begin{split}
    \langle W|\left(T_{12}(u)+T_{23}(u)\right) & = \langle W|\left(T_{12}\left(-u-\frac{i}{2} \right)+T_{23}\left(-u-\frac{i}{2} \right) \right)\,, \\ 
    \langle W|T_{11}(u) & = \langle W|T_{11}\left(-u-\frac{i}{2} \right)\,, \\ 
    \langle W|T_{21}(u) & = \langle W|T_{21}\left(-u-\frac{i}{2} \right)\,, \\ 
    \langle W|T_{31}(u) & = \langle W|T_{31}\left(-u-\frac{i}{2} \right)\,. 
\end{split}
\end{equation}

\paragraph{Invariance under global rotations}

We will now show that $\langle W|$ is annihilated by all global Lie algebra generators and as a consequence is invariant under global rotations. The global Lie algebra generators $\lE_{ij}$ are encoded in the large-$u$ asymptotics of the monodromy matrix elements $T_{ji}$. By explicit construction of $T_{ij}(u)$ from Lax operators we have 
\begin{equation}
    T_{ij}(u) = \left(u+\frac{i}{4}\right)^{2L}\delta_{ij} + i\, \left(u+\frac{i}{4}\right)^{2L-1}\left(\lE_{ji}-\frac{L}{2}\delta_{ij}\right) + \lO\left(u^{2L-1}\right)
\end{equation}
where the global generators are defined as 
\begin{equation}
    \lE_{ij} = \displaystyle \sum_{\alpha=1}^L \bbE_{ij}^{(2\alpha)}+ \bar{\bbE}_{ij}^{(2\alpha-1)}
\end{equation}
where the superscript $(\alpha)$ denotes which site of the chain the generators are acting on. The parity relations \eqref{eqn:newparity} then imply that $\langle W|$ is annihilated by $\lE_{21}+\lE_{32}$, $\lE_{12}$ and $\lE_{32}$ and that it is an eigenstate of $\lE_{11}$ with eigenvalue $\frac{L}{2}$. By applying the usual Lie algebra commutation relations, it follows quickly that $\langle W|$ is annihilated by all off-diagonal $\mathcal{E}_{ij}$, i.e. it is a singlet state invariant under any global rotation. It then follows that $\langle W|$ takes the same form for any choice of twist $\Lambda$, as these are all related to the companion twist by global rotation. 

\subsection{$KT$-relation}

From \eqref{eqn:newparity} we can exploit the commutation relations between $T_{ij}$ to obtain a reflection invariance property for all generators:
\begin{equation}\label{eqn:WTprop}
    \langle W|T_{ij}(u) = \langle W|T_{ij}\left(-u-\frac{i}{2} \right)\,.
\end{equation}
This relation allows us to embed the state $\langle W|$ into the well-studied framework of integrable two-site boundary states. The key relation of such states is the so-called $KT$-relation \cite{Gombor:2021uxz,Gombor:2021hmj,Gombor:2023bez,Gombor:2022deb}, which involves an invertible matrix $K(u)$ and a state $\langle B|$ and reads 
\begin{equation}\label{eqn:KT}
    K(u)\langle B|T(u) = \langle B|T\left(-u-\frac{i}{2}\right)K(u)\,.
\end{equation}
The matrix $K(u)$ is acting in the auxiliary space of the spin chain and $\langle B|$ is in the physical (dual) Hilbert space. Hence this equation actually defines a family of relations as we can project onto different components of the auxiliary space. Any pair $K(u)$ and $\langle B|$ which solve the $KT$ relation specify an integrable boundary state, and clearly our relation \eqref{eqn:WTprop} corresponds to $K=1$.

\subsection{Explicit Realisation}

We will now give an explicit realisation of $\langle W|$ in the standard spin chain basis. Let us look for a solution in a factorised ``two-site" form
\begin{equation}\label{eqn:Wfact}
    \langle W| = \langle W_2| \otimes \dots \otimes \langle W_2|
\end{equation}
and require that $\langle W_2|$ satisfies \eqref{eqn:WTprop} for a spin chain with two sites, i.e. 
\begin{equation}
    \displaystyle \sum_k\langle W_2| \lL_{ik}(u-\theta)\bar{\lL}_{kj}(u+\theta) = \displaystyle \sum_k \langle W_2| \lL_{ik}\left(-u-\frac{i}{2}-\theta\right)\bar{\lL}_{kj}\left(-u-\frac{i}{2}+\theta\right)\,.
\end{equation}
By using the explicit expression of the Lax operators, it is easy to deduce that 
\begin{equation}
    \langle W_2| = \sfe^i \otimes \bar{\sfe}^i
\end{equation}
is the unique solution up to an overall normalisation, where $\sfe^i$ and $\bar{\sfe}^i$ are related to the basis vectors $\sfe_i$ and $\bar{\sfe}_i$ by the orthogonality property 
\begin{equation}
    \sfe^i \sfe_j = \delta^i_{j},\quad \bar{\sfe}^i \bar{\sfe}_j = \delta^i_{j}\,.
\end{equation}
By the product structure of the monodromy matrix, it immediately follows that $\langle W|$ defined by \eqref{eqn:Wfact} satisfies \eqref{eqn:WTprop}, and is explicitly given by 
\begin{equation}\label{eqn:Wexplicit}
    \langle W| = \sfe^{i_1} \otimes \bar{\sfe}^{i_1} \otimes \dots \otimes \sfe^{i_L} \otimes \bar{\sfe}^{i_L}\,.
\end{equation}

So far all we have done is construct \textit{some} solution to \eqref{eqn:WTprop} -- we have no clear indication that it is the one defined via FSoV. However, Gombor has shown that for any integrable boundary state $\langle B|$ satisfying the $KT$-relation the overlaps of $\langle B|$ with any transfer matrix eigenstate is completely fixed (up to normalisation of $\langle B|$) by the property that $\langle B|$ satisfies the $KT$-relation for some $K(u)$ \cite{Gombor:2021hmj,Gombor:2023bez}. Since transfer matrix eigenstates form a basis, their overlaps with $\langle B|$ completely fix $\langle B|$. Hence, for a given $K(u)$, there is a unique up to normalisation $\langle B|$ which solve the $KT$-relation. Hence, $\langle W|$ defined via FSoV is precisely \eqref{eqn:Wexplicit} (up to overall normalisation).

\section{Decoding SoV Representation of Determinants}\la{app:sigmification}

In the derivations in this appendix we will be heavily relying on the following equation, which is a generalisation of the relation used in \cite{Gromov:2020fwh,Gromov:2022waj} for determinants
\beq\la{relation}
\det_{(a,\alpha),(b,\beta)}
H_{\alpha,a,\beta,b} G_{\alpha,a,b}=
\sum_{\sigma}{\rm sig}_\sigma
\prod_{a}
\left(
\left[\prod_{\alpha}
G_{\alpha,a,\sigma_{\alpha,a}}\right]
\left[\det_{(\alpha,b)\in \sigma^{-1}(a),\beta}
H_{\alpha,b,\beta, a}\right]
\right)\;.
\eeq
For the purpose of this paper $\alpha,\beta$ will be running either $1,\dots,K$ where $K=L$ or $K=2L$ depending on the particular case, whereas $a,\;b$ can take two values $1,2$.

The equation extracts the part of the expression from under the determinant which does not depend on $\beta$. If it was also independent of $b$ we could remove it as a constant prefactor, however, there is some combinatorics to take into account for the case when $G$ has a nontrivial $b$ dependence. The result in this case is a sum over permutations $\sigma$ of the sequence of $2K$ numbers $1,2,1,2,\dots$.
Given such a permutation $\sigma_{\alpha,a}$ denotes the number in the sequence at the position $2(\alpha-1)+a$, whereas the inverse $\sigma^{-1}(b)$ gives a set of all pairs $(\alpha,a)$ for which 
$\sigma_{\alpha,a}=b$.

In order to further clarify the notations we included the below {\it Mathematica} code, which also provides a test of this relation for $K=3$. First we define some useful helper functions:
\begin{mathematica}
K = 3;
(*canonical order*)
sig0 := Table[Table[a, {a, 2}], {al, K}] // Flatten;
xs := Flatten[Table[x[al, a], {al, K}, {a, 2}]];
(*Subset of x\[CloseCurlyQuote]s with given value of sigma*)
xsinv[z_, sig_] := xs[[(Position[sig, z] // Flatten)]];
(*Signature of a permutation*)
SGN[sig_] := Signature[Flatten[Table[xsinv[z, sig], {z, 2}]]];
\end{mathematica}
Then we code and compare the r.h.s. and l.h.s. of \eq{relation} 
\begin{mathematica}
(*LHS*)
LHS = Table[{al, a} = List @@ aa; {be, b} = (List @@ bb); 
     H[al, a, be, b] G[al, a, b], {aa, xs}, {bb, 
      xs}] // Det // ExpandAll;

(*RHS*)
sigmas = Permutations[sig0];
RHS = Sum[SGN[sigma0]/SGN[sig0] Product[
   Product[G[al, a, Sequence @@ sigma0[[a + (al - 1) K]]], {al, L}]
   (Table[{al, b} = List @@ ab; 
    H[al, b, be, a], {ab, xsinv[a, sigma0]}, {be, L}] // Det)
      , {a, 2}] , {sigma0, sigmas}] // Expand;
(*check they give the same*)
LHS - RHS    
\end{mathematica}
In the next section we use this relation to derive the SoV form of various determinants giving overlaps in the main text.

\subsection{Derivation of the SoV Measure for $\mathfrak{su}(3)$}
First we consider the expression for the scalar product of transfer matrix eigenstates \eq{eqn:spdet}. We will consider the determinant in the r.h.s. of \eq{eqn:spdet}, which we denote as $\det$ for simplicity
\begin{equation}\label{eqn:spdet3}
    {\det}=
   \det_{(\alpha,a),(s,p)} \left(
    \begin{array}{c}
        \bl \bQ_1^A\; D^{\tfrac{s}2} \;u^{p-1}\; D^{\tfrac{s}2}\; \bQ_B^a\br_\alpha   \\
        \bl \bQ_a^A\; D^{\tfrac{s}2} \;u^{p-1}\; D^{\tfrac{s}2}\;\bQ^1_B\br_{\bar\alpha}  
    \end{array}
    \right)\;.
\end{equation}
Below however, we have found it more convenient to label the Q-functions as $\bQ_1$ and $\bQ_2$ instead of $\bQ_1$ and $\bQ_3$, as well as the index $b=1,2$ instead of $s=\pm 1$, in order for our expressions to be more in line with previous papers \cite{Gromov:2020fwh,Gromov:2022waj} for readers familiar with those works. At the end, all Q-functions will drop out and the objects we compute are invariant under these choices. 

Firstly, we use the definition of the bracket \eq{brdef} to pull out the factors independent of $\bQ_1^A$ and $\bQ^1_B$
\beq\la{d2intsL}
\det =
 \int t(\{u_{\alpha,a}\},\{u_{\bar\alpha,a}\})
 \prod_{\alpha=1}^L\prod_{a=1}^2\frac{{\rm d}{u}_{\alpha,a}}{2\pi i}
 \frac{{\rm d}{u}_{\bar\alpha,a}}{2\pi i}
\bQ_1^A({u}_{\alpha,a})\bQ_B^1({u}_{\bar\alpha,a})\rho_\alpha({u}_{\alpha,a})\rho_{\bar\alpha}({u}_{\bar\alpha,a}+\tfrac{i}{2})
\eeq
where
\beq
t(\{u_{\alpha,a}\},\{u_{\bar\alpha,a}\})\equiv \det_{(a,m),(b,n)}
\left[
\bea{c}
\;\(u_{\alpha,a}+i \frac{3-2b}{4}\)^{n}\;  \bQ_B^{a}\(u_{\alpha,a}+i \frac{3-2b}{2}\)\\
\;\(u_{\bar\alpha,a}-i \frac{3-2b}{4}\)^{n}\;  \bQ_a^{A}\(u_{\bar\alpha,a}-i \frac{3-2b}{2}\) 
\eea
\right]\;.
\eeq
where we introduce $m=\alpha\cup\bar\alpha$, a $2L$ dimensional index, which joins together two $L$ dimensional indexes $\alpha$ and $\bar\alpha$, labeling the rows of the matrix in the r.h.s. together with the $2$ dimensional index $a$. Next, we use the relation \eq{relation} by identifying
\beq
G_{m,a,b}=\left(
\bea{c}
\bQ_B^{a}\(u_{\alpha,a}+i \frac{3-2b}{2}\)\\
\bQ_a^{A}\(u_{\bar\alpha,a}-i \frac{3-2b}{2}\) 
\eea
\right)\;\;,\;\;H_{m,a,n,b}=
\left[
\bea{c}
\;\(u_{\alpha,a}+i \frac{3-2b}{4}\)^{n-1}\\
\;\(u_{\bar\alpha,a}-i \frac{3-2b}{4}\)^{n-1}
\eea
\right]\;.
\eeq
Applying \eq{relation} we get
\beqa\nn
t(\{u_{\alpha,a}\},\{u_{\bar\alpha,a}\}) &=&
\sum_\sigma {\rm sig}_\sigma
\prod_{a=1,2}
\left(
\left[
\prod_{\alpha}
\bQ_B^{a}\left(u_{\alpha,a}+i s_{\alpha,a}+\tfrac{i}{2}\right)
\prod_{\bar\alpha}
\bQ_a^{A}\left(u_{\bar\alpha,a}-i \bar s_{\alpha,a}-\tfrac{i}{2}\right)
\right]
\right)\\
&&\prod_{a=1,2}
\det_{(m,b)\in \sigma^{-1}(a),n}
\left[
\bea{c}
\;\(u_{\alpha,b}+ i\tfrac{3-2a}{4}\)^{n-1}\\
\;\(u_{\bar\alpha,b}- i\tfrac{3-2a}{4}\)^{n-1}
\eea
\right]\;.
\eeqa
Next we notice that the determinant can be written in terms of the Vandermond. Furthermore, using that the integration measure is symmetric under $u_{m,1}\leftrightarrow u_{m,2}$
we get 
\beqa\nn
\det=\sum_\sigma {\rm sig}_\sigma
\int&&\prod_{\alpha=1}^L\prod_{a=1}^2\frac{{\rm d}{u}_{\alpha,a}}{2\pi i}
 \frac{{\rm d}{u}_{\bar\alpha,a}}{2\pi i}
\rho_\alpha({u}_{\alpha,a})\rho_{\bar\alpha}({u}_{\bar\alpha,a}+\tfrac{i}{2})
\bQ_1^A({u}_{\alpha,a})\bQ_B^1({u}_{\bar\alpha,a})\\
&&\prod_{\alpha}
\bQ_B^{[1}\left(u_{\alpha,1}+i s_{\alpha,1}+\tfrac{i}{2}\right)
\bQ_B^{2]}\left(u_{\alpha,2}+i s_{\alpha,2}+\tfrac{i}{2}\right)
\\
 &&
\prod_{\bar\alpha}
\bQ_{[1}^{A}\left(u_{\bar\alpha,1}-i s_{\bar\alpha,1}-\tfrac{i}{2}\right)
\bQ_{2]}^{A}\left(u_{\bar\alpha,2}-i s_{\bar\alpha,2}-\tfrac{i}{2}\right)
\\
&&\prod_{a}
\Delta\left(\{u_{m,b}+ i\tfrac{3-2a}{4}S_{m}\}_{(m,b)\in \sigma^{-1}(a)}\right)
\eeqa
where $S_{m}=1$ for $m=1,\dots,L$ and $S_{m}=-1$ for $m>L$ and $s_{\alpha,a}=1-\sigma_{\alpha,a}$.

Next, we compute the integral by residues. 
We use the key property of 
$\rho_\alpha(u)$ that it has zeroes at all points where poles in $\bQ$ can appear except for $u=\theta_{\alpha}\pm\tfrac{i}{2}$. Similarly in
$\rho_{\bar\alpha}(u+\tfrac i2)$ the missing zeroes are at $-\theta_{\bar\alpha}\pm\tfrac{i}{2}$ so the poles in the expression are situated at
\beq
u_{\alpha,a}=\theta_\alpha+i n_{\alpha,a}-\tfrac{i}{2}\;\;,\;\;
u_{\bar\alpha,a}=-\theta_\alpha+i \bm_{\bar\alpha,a}-\tfrac{i}{2}\;\;,\;\;n_{\alpha,a},\;m_{\bar\alpha,a}=0,1\;.
\eeq
We see that at the poles locations the arguments of $\bQ$ became those of $\svx$ and $\svy$.
However for the given $\sigma$ the SoV labels $n_{\alpha},\;\bar n_{\bar\alpha}$ and $m_{\alpha},\;\bar m_{\bar\alpha}$,
parametrising $\svx$ and $\svy$ \eq{eq:xinn}, \eq{ym} have to be related as follows
\beq\la{nton}
\sigma_{\alpha,a} =  n_{\alpha,a}-m_{\alpha,a}+a\;\;,\;\;\sigma_{\bar\alpha,a}= \bn_{\bar\alpha,a}-\bm_{\bar\alpha,a}+3-a\;.
\eeq
This can be written in a compact form as $\svy=\svx-i\sigma+\tfrac{3i}{2}$ so that finally we arrive to the following compact expression
\beqa\la{detfin}
\det=&&
\sum_\sigma{\rm sig}_\sigma
\sum_{\svx}
\;
\left.
\langle {\svx}|\Psi_A\rangle
\langle \Psi_B|{\svy}\rangle
\prod_{a}\Delta\(\frac{{\svx}+{\svy}}{2}:\sigma=a\)\right|_{{\svy} = {\svx}-i\sigma+\tfrac{3i}{2}}
\eeqa
From this expression it is now easy to extract the SoV measure $\langle \svy|\svx\rangle$ by comparing \eq{detfin}  with
\beq
\det=
{\cal N}\langle \Psi_B|\Psi_A\rangle = 
{\cal N}\sum_{\svx,\svy}\langle \Psi_B|{\svy}\rangle
\langle {\svy}|{\svx}\rangle
\langle {\svx}|\Psi_A\rangle   \;.
\eeq
This expression allows also to determine the normalization coefficient ${\cal N}$ from the requirement that $\langle 0|0\rangle$ for the overlap between two SoV vacua. 
In section \ref{app:code} we give a compact {\it Mathematica} code for $\langle \svx|\svy\rangle$, which should further clarify our notations.

\subsection{Derivation of the Expression for the Boundary State in SoV Basis}
Here we repeat the same procedure for the determinant appearing in \eq{eqn:spdet2}. We duplicate it here for convenience
\label{app:Wx}
\begin{equation}\label{eqn:spdet2startW}
    d_L=\displaystyle\det_{(a,\alpha),(s,\beta)}         \bl \bQ_1(u)\; D^{\tfrac{s}2} \;u^{2\beta-1}\; D^{\tfrac{s}2}\; \bQ_a(-u)\br_{\alpha} \;.
\end{equation}
Again we pull out the common factors and the integrals out of the determinant
\beq
{\det}_+ =
 \int t(\{u_{\alpha,a}\})
 \prod_{\alpha=1}^L\prod_{a=1}^2\frac{{\rm d}{u}_{\alpha,a}}{2\pi i}
\bQ_1({u}_{\alpha,a})
\rho_\alpha({u}_{\alpha,a})
\eeq
where
\beq\la{tdet}
t(\{u_{\alpha,a}\})\equiv \det_{(a,\alpha),(b,\beta)}
\left[
\;\(u_{\alpha,a}+i \frac{3-2b}{4}\)^{2\beta-1}\;  \bQ_{a}\(-u_{\alpha,a}-i \frac{3-2b}{2}\)
\right]\;.
\eeq
By identifying
\beq
G_{\alpha,a,b}=\left(
\bQ_{a}\(-u_{\alpha,a}-i \frac{3-2b}{2}\)
\right)\;\;,\;\;H_{\alpha,a,\beta,b}=
\left[
\;\(u_{\alpha,a}+i \frac{3-2b}{4}\)^{2\beta-1}
\right]
\eeq
and applying \eq{relation} we get
\beqa\nn
t &=&
\sum_\sigma {\rm sig}_\sigma
\prod_{a=1,2}
\prod_{\alpha}
\bQ_B^{a}\left(u_{\alpha,a}+i s_{\alpha,a}+\tfrac{i}{2}\right)
\prod_{a=1,2}
\det_{(\alpha,b)\in \sigma^{-1}(a),\beta}
\left[
\;\(u_{\alpha,b}+ i\tfrac{3-2a}{4}\)^{2\beta-1}
\right]\;.
\eeqa
Next computing the determinant and using that the integration measure is symmetric 
up to a relabeling of the integration variables we get
\beqa\nn
{\det}_+=\sum_\sigma {\rm sig}_\sigma
\int&&\prod_{\alpha=1}^L\prod_{a=1}^2\frac{{\rm d}{u}_{\alpha,a}}{2\pi i}
\rho_\alpha({u}_{\alpha,a})
\bQ_1^A({u}_{\alpha,a})\\
&&\prod_{\alpha}
\bQ_{[1}\left(-u_{\alpha,1}-i s_{\alpha,1}-\tfrac{i}{2}\right)
\bQ_{2]}\left(-u_{\alpha,2}-i s_{\alpha,2}-\tfrac{i}{2}\right)
\\
&&\prod_{a}
\Delta_+\left(\{u_{\alpha,b}+ i\tfrac{3-2a}{4}\}_{(\alpha,b)\in \sigma^{-1}(a)}\right)
\eeqa
where $\Delta_+$ is defined as
\beq
\Delta_+(\{A_i\})=\prod_{i} A_i \prod_{i<j}(A_i^2-A_j^2)\;.
\eeq
Next we compute the integral by residues. 
For given $\alpha$ the only residues are at
\beq
u_{\alpha,a}=\theta_\alpha+i n_{\alpha,a}-\tfrac{i}{2}\;\;,\;\;n_{\alpha,a}=0,1\;.
\eeq
We see that the argument of $\bQ$'s are the SoV variables $\svx$ which are convenient to split into halves $\svx$ and $\bar{\svx}$.
We see that in fact ${\svx}$ and $\bar{\svx}$ are not independent but are related due to the following constraint
\beq\la{ntomW}
\sigma_{\alpha,a} =  n_{\alpha,a}+\bn_{\alpha,a}+2-a\quad{\rm or}\quad{\svx}_{\alpha,a}+\bar{\svx}_{\alpha,a}=i\sigma_{\alpha,a}-\frac{3i}{2}\;. 
\eeq
Finally we get the following compact expression:
\beqa
{\det}_+=&&
\sum_\sigma{\rm sig}_\sigma\sum_{\svx}\;
\left.
\langle {\svx}|\Psi\rangle
\prod_{a}\Delta_+\(\left\{\frac{{\svx}_{\alpha,b}-\bar{\svx}_{\alpha,b}}{2}\right\}_
{({\alpha,b})\in \sigma^{-1}(a)}
\)\right|_{\bar{\svx}_{\alpha,a}=i\sigma_{\alpha,a}-{\bf x}_{\alpha,a}-\tfrac{3i}{2}}\;.
\eeqa
By comparing with \eq{detdec} one can easily extract $\langle W|x\rangle = {\cal W}_x$.
In the next section we give a compact {\it Mathematica} code, computing ${\cal W}_x$ to help the reader to further clarify our notations. 
\subsection{{\it Mathematica} code for $\langle\svy|\svx\rangle$ and $\langle W|\svx\rangle$}\la{app:code}
In this section we give our implementation for the SoV measure $\langle\svy|\svx\rangle$ and for the Boundary state in SoV basis $\langle W|\svx\rangle$. We begin by introducing a few helper functions:
\begin{mathematica}
Nc = 3;
(*Subset of x\[CloseCurlyQuote]s with given value of sigma*)
xsinv[L_, z_, sig_] := xs[L][[(Position[sig, z] // Flatten)]];
(*Signature of a permutation*)
SGN[L_,sig_] := Signature[Flatten[Table[xsinv[L,z, sig], {z, Nc - 1}]]];
(*Trivial permutation*)
sig0[L_] := Table[Table[a, {a, Nc - 1}], {al, L}] // Flatten;
xs[L_] := Flatten[Table[X[al, a], {al, 2 L}, {a, Nc - 1}]];
(*Subset of x\[CloseCurlyQuote]s with given value of sigma*)
xsinv[L_, z_, sig_] := xs[L][[(Position[sig, z] // Flatten)]];
(*Vandermond determinants*)
Delta[lst_] := Product[lst[[i]] - lst[[j]], {i, Length[lst]}, {j,i+1,Length[lst]}];
DeltaP[lst_] := Product[lst[[i]], {i, Length[lst]}] Product[lst[[i]]^2 - lst[[j]]^2,
   {i, Length[lst]}, {j, i + 1, Length[lst]}];
(*Creates all permutations with fixed \alpha's*)
MyPerm[L_] := Flatten[Table[Flatten@Table[p[al], {al, L}], Evaluate[
     Sequence @@ Table[{p[al], Permutations[
         Take[#, {(Nc - 1) (al - 1) + 1, (Nc - 1) al}]]}, {al, L}]]], L - 1] &
\end{mathematica}
The code for the measure:
\begin{mathematica}
(*Normalization constant*)
goN[L_] := Delta[Table[t[a] - I/4, {a, L}]~Join~Table[-t[a] - 3 I/4, {a, L}]] *
           Delta[Table[-t[a] - I/4, {a, L}]~Join~Table[t[a] - 3 I/4, {a, L}]];
Sm[L_] := Table[If[i <= 2 L, 1, -1], {i, 4 L}];
yx[L_, ms_, msb_, ns_, nsb_] := 1/goN[L] If[
    (*check SoV charges are the same*)
    Total[ms] + Total[msb] == Total[ns] + Total[nsb],
    sig = (Join[ns, nsb] - Join[ms, msb]) + 3/2 + 
      Sm[L] (sig0[2 L] - 3/2);
    If[(*check if sig is a valid permutation*)
     Union[Tally[sig]] == Union[Tally[sig0[2 L]]],
     Clear[k, m, x, y];
     k[al_, a_] := (ns~Join~nsb)[[(al - 1) (Nc - 1) + a]];
     m[al_, a_] := (ms~Join~msb)[[(al - 1) (Nc - 1) + a]];
     Do[
      x[al, a] = t[al] + I k[al, a] - I/2;
      x[al + L, a] = -t[al] + I (k[al + L, a] + 1 - a);
      y[al, a] = t[al] + I (m[al, a] + 1 - a);
      y[al + L, a] = -t[al] + I m[al + L, a] - I/2;
      , {al, L}, {a, Nc - 1}];
     (*the main formula*)
     SGN[L,sig]/SGN[L,sig0[2 L]]*
      Product[Delta[
        xsinv[L, z, sig] /. X[p_, b_] :> (x[p, b] + y[p, b])/2], {z, 
        Nc - 1}], 0], 0] /. t -> \[Theta]
\end{mathematica}
The code for the boundary state in the SoV basis:
\begin{mathematica}
Wx[L_, ns_, nsbar_] := If[(*check SoV charges are the same*)
    Total[ns] + Total[nsbar] == 2 L,
    Sum[
     sig = (ks + nsbar) + (2 - sig0[L]);
     If[(*check if sig is a valid permutation*)
      Union[Tally[sig]] == Union[Tally[sig0[L]]],
      Clear[n, nb, x, xb];
      n[al_, a_] := ks[[(al - 1) (Nc - 1) + a]];
      nb[al_, a_] := nsbar[[(al - 1) (Nc - 1) + a]];
      Do[
       x[al, a] = \[Theta][al] + I n[al, a] - I/2;
       xb[al, a] = -\[Theta][al] + I (nb[al, a] - a + 1);
       , {al, L}, {a, Nc - 1}];
      (*the main formula*)
      SGN[L,sig]/SGN[L,sig0[L]]*
       Product[DeltaP[
         xsinv[L, z, sig] /. X[p_, b_] :> (x[p, b] - xb[p, b])/2], {z,
          Nc - 1}], 0], {ks, MyPerm[L][ns]}], 0] /. t -> \[Theta]
\end{mathematica}
For example by running
\begin{mathematica}
Wx[2, {1, 1, 0, 0}, {0, 0, 1, 1}] /. \[Theta] -> t // Factor    
\end{mathematica}
you should get
\begin{mathematica}
((4 t[1] + I) (4 t[1] + 3 I) (t[1] - t[2] + I)^2 (2 t[1] + 2 t[2] - 
   I) (2 t[1] + 2 t[2] + I) (4 t[2] - I) (4 t[2] - 3 I))/1024
\end{mathematica}

\section{Generalisation to $\mathfrak{su}(N)$}\la{app:su_n_generalization}

In this appendix, we generalise the $\mathfrak{su}(3)$ example of the main text to $\mathfrak{su}(N)$. As will be clear, the generalisation is relatively straightforward and all arguments follow closely those already presented. The final formulas are identical, with the only difference being the extension of the range of indices. We will therefore be brief and omit most details. We will only present formulas for the scalar product $\langle \Psi_A|\Psi_B\rangle$ and the overlap $\braket{W}{\Psi}=\langle \Psi|W\rangle$ -- insertions can also be derived using the methods of the main text. 

\subsection{Q-system and Baxter Equations}
The Q-system for $\mathfrak{su}(N)$ is built from functions $Q_{A}$ where $A$ is a multi-index that contains lower case Latin indices with range $a=1,\dots,N$. Upper index Q-functions are defined in terms of lower index Q-functions through
\begin{equation}
    Q^{A}\, \propto\, \frac{1}{\prod_{b=1}^{N} \lambda_{b}^{\ii u}}\sum_{B} \frac{\epsilon^{AB}}{|B|!}Q_{B}
\end{equation}
where the proportionality factor ensures that $Q^a \sim \lambda_a^{-i u}u^M$ for some $M$. We consider spin chains with Hilbert space of type $\bigotimes^{L} \mathbf{N} \otimes \overline{\mathbf{N}}$, thus $Q_{a}$ and $Q^{a}$ satisfy the same QQ-relations as in the main text
\begin{equation}
    \Wr(Q_{a},Q_{b}) = Q_{\theta} \, Q_{ab}\;,
    \quad
    \Wr(Q^{a},Q^{b}) = Q^{ab}Q_{\btheta}\;.
\end{equation}
and source term are also the same as before
\begin{equation}
    Q_{\theta}=\prod_{\alpha=1}^L (u-\theta_{\alpha})\;,\quad Q_{\btheta} = \prod_{\alpha=1}^L(u-\btheta_{\alpha})\;.
\end{equation}
The source term does not make an appearance in the remaining QQ-relations which are written as
\begin{equation}
    \Wr(Q_{Aa},Q_{Ab}) = Q_{Aab}Q_{A}\;, \quad |A|\neq 0,N-2\;.
\end{equation}
The Baxter finite-difference operator $\lO$ is given by
\begin{equation}\label{BaxtersuN}
\begin{split}
    &\mathcal{O} = \\
    &\frac{1}{Q^+_{\theta}Q^-_{\theta}}\left[\chi_{+N}{Q_{{\theta}}^{-}}D^{N}{Q_{\Bar{\theta}}^{+}}
    +\displaystyle \sum_{\beta=1}^{N-1}\tau_{N-2r} D^{N-2r}   
    +(-1)^N\chi_{-N}{Q^{+}_{\theta}}D^{-N}{Q^{-}_{\bar\theta}}\right]\frac{1}{Q^{+}_{\Bar{\theta}}Q^{-}_{\Bar{\theta}}}\,.
\end{split}
\end{equation}
This operator satisfies
\begin{equation}
     Q_a \overleftarrow{\mathcal{O}}= 0\;,\quad \overrightarrow{\mathcal{O}} Q_a = 0\;,
\end{equation}
and for convenience we again introduce the rescaled Q-functions 
\begin{equation}
    \bQ_a = \frac{Q_a}{Q_\theta^+ Q_\theta^-}\;,\quad \bQ^a = \frac{Q^a}{Q_{\bar\theta}^+ Q_{\bar\theta}^-}\;.
\end{equation}
Then, using the same definitions of the bracket as in the $\mathfrak{su}(3)$ case we have
we have the identity
\begin{equation}
    \bl f \overrightarrow{\mathcal{O}} g \br= \bl (f \overleftarrow{\mathcal{O}} g)^{[N]} \br\,.
\end{equation}

The transfer matrix corresponding to the usual trace of the monodromy matrix is $\tau_{N-2}$, with $\tau_{N-2r}$ corresponding to the $r$-th fused anti-symmetric representation, i.e. the vertical Young diagram with $r$ boxes. They admit the following expansion into a complete basis of integrals of motion $I_{r,n}$
\begin{equation}
    \tau_r(u) = \chi_r \left(u+\frac{i\,r}{4} \right)^{2L}+\displaystyle \sum_{n=0}^{2L-1} \left(u+\frac{i\,r}{4} \right)^{n} I_{r,n}\,.
\end{equation}

\subsection{The Scalar Product and Boundary Overlap}
Repeating the arguments of Section~\ref{sec:crash_course_fsov} we find the scalar product. It is given by
\begin{equation}\label{eqn:spdetSUN}
    \langle \Psi_B|\Psi_A\rangle = \frac{1}{\cal N}\displaystyle\det_{(a,\alpha),(s,p)} \left(
    \begin{array}{c}
        \bl \bQ_1^A\; D^{\tfrac{s}2} \;u^{p-1}\; D^{\tfrac{s}2}\; \bQ_B^a\br_\alpha   \\
        \bl \bQ_a^A\; D^{\tfrac{s}2} \;u^{p-1}\; D^{\tfrac{s}2}\;\bQ^1_B\br_{\bar\alpha}  
    \end{array}
    \right)
\end{equation}
with the only difference from the $\mathfrak{su}(3)$ case being that now the index $s$ runs over $N-2,\dots,2-N$ and the index $a$ runs over $1,\dots,N-1$. Note that this differs from the main text where we chose to use $\bQ_1$ and $\bQ_3$ instead of $\bQ_1$ and $\bQ_2$ -- it simply amounts to a different choice of highest-weight state. Finally, the measures are given by
\begin{equation}
\begin{split}
&\rho_\alpha(u)=\frac{1}{{\cal R}_\alpha}\prod_{\beta\neq \alpha}^L (1+e^{2\pi (u-\theta_\beta)})
\prod_{\beta=1}^L (1+e^{-2\pi (u+\theta_\beta+\ii \tfrac{(N-2)}{2})})\;,\\
&\rho_{\bar\alpha}(u)=\rho_{\alpha}(-u+i\tfrac{N-2}{2})\;.
\end{split}
\end{equation}

\paragraph{Reflection.}
The integrable boundary state $\langle W|$ is now defined as the singlet of $\mathfrak{su}(N)$ satisfying the reflection property
\begin{equation}
    \langle W|\hat{\tau}_r\left(u \right) = \langle W|\hat{\tau}_r\left(-u-\frac{i\,r}{2} \right),\quad r=N-2,\dots,-N+2\,.
\end{equation}
For states $|\Psi\rangle$ with non-zero overlap with $\langle W|$ the Q-functions must satisfy the symmetry property $\bQ_a(-u)=\bQ^a(u)$. Finally, we can deduce the overlap $\langle W|\Psi\rangle$ from FSoV and it is given by
\begin{equation}
    \langle W|\Psi\rangle = \displaystyle\det_{(a,\alpha),(s,\beta)}         \bl \bQ_1(u)\; D^{\tfrac{s}2} \;u^{2\beta-1}\; D^{\tfrac{s}2}\; \bQ_a(-u)\br_{\alpha}\;,
\end{equation}
again exactly the same as in the $\mathfrak{su}(3)$ case up to the range of indices.

\section{Constructing a Basis of Integrable Boundary States}{\label{app:basis_integrable_states}

We will now show that the set of vectors \label{Bdystate} are linear independent. To do this, we will explicitly show that they are linear independent in the limit where the inhomogeneities $\theta_\alpha$ become relatively large $|\theta_\alpha-\theta_\beta|\gg 1$. Since the vectors $\langle B_n|$ are polynomial functions of the inhomogeneities, up to normalisation, proving linear independence in this limit establishes it for all values apart from a subset of measure zero. 

The basis is constructed by repeated action of the following ``raising operators"
\begin{equation}
    \frac{\hat{\tau}_+(\theta_\alpha+i/2)}{Q_\theta^{[2]}(\theta_\alpha)Q_{\bar\theta}^{[3]}(\theta_\alpha)}\,.
\end{equation}
In the limit $|\theta_\alpha -\theta_\beta|\gg 1$, the transfer matrix degenerates 
\begin{equation}
   \frac{\hat{\tau}_+(\theta_\alpha+i/2)}{Q_\theta^{[2]}(\theta_\alpha)Q_{\bar\theta}^{[3]}(\theta_\alpha)} \sim \mathcal{L}^{(\alpha)}_{ij}(i/2)\Lambda_{ji} = \Lambda^{(\alpha)}
\end{equation}
where we remind that $\Lambda$ is the twist matrix, and $\Lambda^{(\alpha)}$ denotes the twist matrix acting non-trivially on the $\alpha$-th site of the chain and we have used the explicit form of the Lax matrix in terms of basis matrices of $GL(3)$. Hence, in this limit the vectors $\langle B_n|$ degenerate to 
\begin{equation}
    \langle B_n| \rightarrow \left(\langle W_2|\Lambda^{n_1}\right)\otimes \dots \otimes \left(\langle W_2|\Lambda^{n_L}\right)
\end{equation}
where $\langle W_2|$ denotes the $2$-site state composing the full $\langle W|$. Since the states are factorised, to prove linear independence it remains to prove that each of the vectors $\langle W|\Lambda^n$ are linear independent for $n=0,1,2$. It is trivial to show they are once the matrix $\Lambda$ is in companion form, see for example \cite{Maillet:2018bim}.

\paragraph{Charge.} There is a charge operator which one can construct to measure the number of excitations a given state $\langle B_n|$ has above $\langle W|$, at least in the limit where $\chi_+=\chi_-=\chi_{-3}=0$ and all inhomogeneities are largely separated. In this limit, as discussed above, excitations are created with local operators $\sfe_{21} +\sfe_{32}$, see \eqref{Lambdadefiningbasis}, which act as raising operators for the global Cartan element $\mathcal{E}_{11}-\mathcal{E}_{33}$. In the discussed limit, this operator acts on $\langle B_n|$ with eigenvalue $\sum_{\alpha}n_{\alpha,1}+n_{\alpha,2}$. Hence, the only state which can be a singlet under the global symmetry algebra is $\langle W|$ as all other states have charge where $\mathcal{E}_{11}\neq \mathcal{E}_{33}$.

\section{Character Projection}\la{app:proofs_multiple_insertions}

Let's recall the relation \eqref{beforeproj} which reads, after expanding the integrals of motion linearly in characters (valid in the companion twist frame),
\begin{equation}\label{beforeproj2}
    \sum_{r=\pm 1,3} \chi_r \mel{W}{\hat{I}^{\gamma,(r)}_{s,2\beta-2}}{\Psi} = s\sum_{r=\pm 1,3} \chi_{r} [(s,\beta)\rightarrow \mathcal{O}_{(r)}]\,.
\end{equation}
Our aim is to derive the ``character projected" version of this equality, which reads
\begin{equation}
    \mel{W}{\hat{I}^{\gamma,(r)}_{s,2\beta-2}}{\Psi} = s  [(s,\beta)\rightarrow \mathcal{O}_{(r)}]\,,
\end{equation}
We will not write out all details -- the manipulations are identical to what was carried out in \cite{Gromov:2022waj} as well as in Appendix \ref{app:sigmification}.

The starting point is the sum of determinants on r.h.s. of \eqref{beforeproj2}. Each of these determinants can be expressed as a sum over wave functions in the SoV basis. This allows us to write
\begin{equation}
    s\sum_{r=\pm 1,3} \chi_{r} [(s,\beta)\rightarrow \mathcal{O}_{(r)}] = s\sum_{\svx}\sum_{r=\pm 1,3}\chi_r \mathcal{M}_\svx^{(r),s,\beta} \langle \svx|\Psi_A\rangle\,.
\end{equation}
Now consider the l.h.s. and insert a resolution of the identity $1=\sum_\svx |\svx\rangle \langle \svx|$ giving
\begin{equation}
  \sum_{\svx} \sum_{r=\pm 1,3} \chi_r \langle W|\hat{I}^{\gamma,(r)}_{s,2\beta-2}|\svx\rangle \langle \svx|\Psi\rangle = s\sum_{\svx}\sum_{r=\pm 1,3}\chi_r \mathcal{M}_\svx^{(r),s,\beta} \langle \svx|\Psi_A\rangle\,,
\end{equation}
Since the integral of motion eigenstates $|\Psi\rangle$ form a complete basis of states, we can strip this part off of the previous relation and contract with some fixed $|\svx\rangle$, yielding
\begin{equation}
    \sum_{r=\pm 1,3} \chi_r \langle W|\hat{I}^{\gamma,(r)}_{s,2\beta-2}|\svx\rangle=s\sum_{r=\pm 1,3}\chi_r \mathcal{M}_\svx^{(r),s,\beta}\,.
\end{equation}
The objects $|\svx\rangle$, $\langle W|$, $\hat{I}^{\gamma,(r)}_{s,2\beta-2}$ and $\mathcal{M}_\svx^{(r)}$ are all independent of twist. As a result the only way this relation can be true is if 
\begin{equation}
     \langle W|\hat{I}^{\gamma,(r)}_{s,2\beta-2}|\svx\rangle=s \mathcal{M}_\svx^{(r),s,\beta},\quad r=\pm 1,\pm 3\,.
\end{equation}
Hence, we obtain 
\begin{equation}
    \langle W|\hat{I}^{\gamma,(r)}_{s,2\beta-2}|\Psi\rangle = s[(s,\beta)\rightarrow \mathcal{O}_{(r)}]\,,
\end{equation}
valid for all $r,s=\pm 3\pm 1$, $\beta=1,\dots,L$, and any complex number $\gamma$. The logic for higher insertions is identical.


\bibliographystyle{JHEP}
\bibliography{ref}

\providecommand{\href}[2]{#2}\begingroup\raggedright\begin{thebibliography}{10}

\bibitem{Cavaglia:2018lxi}
A.~Cavagli\`a, N.~Gromov and F.~Levkovich-Maslyuk, \emph{{Quantum spectral
  curve and structure constants in $ \mathcal{N}=4 $ SYM: cusps in the ladder
  limit}}, \href{https://doi.org/10.1007/JHEP10(2018)060}{\emph{JHEP}
  {\bfseries 10} (2018) 060}
  [\href{https://arxiv.org/abs/1802.04237}{{\ttfamily 1802.04237}}].

\bibitem{Cavaglia:2019pow}
A.~Cavagli\`a, N.~Gromov and F.~Levkovich-Maslyuk, \emph{{Separation of
  variables and scalar products at any rank}},
  \href{https://doi.org/10.1007/JHEP09(2019)052}{\emph{JHEP} {\bfseries 09}
  (2019) 052} [\href{https://arxiv.org/abs/1907.03788}{{\ttfamily
  1907.03788}}].

\bibitem{Gromov:2019wmz}
N.~Gromov, F.~Levkovich-Maslyuk, P.~Ryan and D.~Volin, \emph{{Dual Separated
  Variables and Scalar Products}},
  \href{https://doi.org/10.1016/j.physletb.2020.135494}{\emph{Phys. Lett. B}
  {\bfseries 806} (2020) 135494}
  [\href{https://arxiv.org/abs/1910.13442}{{\ttfamily 1910.13442}}].

\bibitem{Gromov:2020fwh}
N.~Gromov, F.~Levkovich-Maslyuk and P.~Ryan, \emph{{Determinant form of
  correlators in high rank integrable spin chains via separation of
  variables}}, \href{https://doi.org/10.1007/JHEP05(2021)169}{\emph{JHEP}
  {\bfseries 05} (2021) 169}
  [\href{https://arxiv.org/abs/2011.08229}{{\ttfamily 2011.08229}}].

\bibitem{Cavaglia:2021mft}
A.~Cavagli\`a, N.~Gromov and F.~Levkovich-Maslyuk, \emph{{Separation of
  variables in AdS/CFT: functional approach for the fishnet CFT}},
  \href{https://doi.org/10.1007/JHEP06(2021)131}{\emph{JHEP} {\bfseries 06}
  (2021) 131} [\href{https://arxiv.org/abs/2103.15800}{{\ttfamily
  2103.15800}}].

\bibitem{Bercini:2022jxo}
C.~Bercini, A.~Homrich and P.~Vieira, \emph{{Structure Constants in
  $\mathcal{N} = 4$ SYM and Separation of Variables}},
  \href{https://arxiv.org/abs/2210.04923}{{\ttfamily 2210.04923}}.

\bibitem{Basso:2022nny}
B.~Basso, A.~Georgoudis and A.~K. Sueiro, \emph{{Structure Constants of Short
  Operators in Planar N=4 Supersymmetric Yang-Mills Theory}},
  \href{https://doi.org/10.1103/PhysRevLett.130.131603}{\emph{Phys. Rev. Lett.}
  {\bfseries 130} (2023) 131603}
  [\href{https://arxiv.org/abs/2207.01315}{{\ttfamily 2207.01315}}].

\bibitem{Gromov:2022waj}
N.~Gromov, N.~Primi and P.~Ryan, \emph{{Form-factors and complete basis of
  observables via separation of variables for higher rank spin chains}},
  \href{https://doi.org/10.1007/JHEP11(2022)039}{\emph{JHEP} {\bfseries 11}
  (2022) 039} [\href{https://arxiv.org/abs/2202.01591}{{\ttfamily
  2202.01591}}].

\bibitem{deLeeuw:2015hxa}
M.~de~Leeuw, C.~Kristjansen and K.~Zarembo, \emph{{One-point Functions in
  Defect CFT and Integrability}},
  \href{https://doi.org/10.1007/JHEP08(2015)098}{\emph{JHEP} {\bfseries 08}
  (2015) 098} [\href{https://arxiv.org/abs/1506.06958}{{\ttfamily
  1506.06958}}].

\bibitem{BuhlMortensen:2015gfd}
I.~Buhl-Mortensen, M.~de~Leeuw, C.~Kristjansen and K.~Zarembo, \emph{{One-point
  Functions in AdS/dCFT from Matrix Product States}},
  \href{https://doi.org/10.1007/JHEP02(2016)052}{\emph{JHEP} {\bfseries 02}
  (2016) 052} [\href{https://arxiv.org/abs/1512.02532}{{\ttfamily
  1512.02532}}].

\bibitem{deLeeuw:2017cop}
M.~de~Leeuw, A.~C. Ipsen, C.~Kristjansen and M.~Wilhelm, \emph{{Introduction to
  integrability and one-point functions in $\mathcal N=$ 4 supersymmetric
  Yang\textendash{}Mills theory and its defect cousin}},
  \href{https://arxiv.org/abs/1708.02525}{{\ttfamily 1708.02525}}.

\bibitem{DeLeeuw:2018cal}
M.~De~Leeuw, C.~Kristjansen and G.~Linardopoulos, \emph{{Scalar one-point
  functions and matrix product states of AdS/dCFT}},
  \href{https://doi.org/10.1016/j.physletb.2018.03.083}{\emph{Phys. Lett. B}
  {\bfseries 781} (2018) 238}
  [\href{https://arxiv.org/abs/1802.01598}{{\ttfamily 1802.01598}}].

\bibitem{deLeeuw:2019usb}
M.~de~Leeuw, \emph{{One-point functions in AdS/dCFT}},
  \href{https://doi.org/10.1088/1751-8121/ab15fb}{\emph{J. Phys. A} {\bfseries
  53} (2020) 283001} [\href{https://arxiv.org/abs/1908.03444}{{\ttfamily
  1908.03444}}].

\bibitem{Linardopoulos:2020jck}
G.~Linardopoulos, \emph{{Solving holographic defects}},
  \href{https://doi.org/10.22323/1.376.0141}{\emph{PoS} {\bfseries CORFU2019}
  (2020) 141} [\href{https://arxiv.org/abs/2005.02117}{{\ttfamily
  2005.02117}}].

\bibitem{Komatsu:2020sup}
S.~Komatsu and Y.~Wang, \emph{{Non-perturbative defect one-point functions in
  planar $\mathcal{N}=4$ super-Yang-Mills}},
  \href{https://doi.org/10.1016/j.nuclphysb.2020.115120}{\emph{Nucl. Phys. B}
  {\bfseries 958} (2020) 115120}
  [\href{https://arxiv.org/abs/2004.09514}{{\ttfamily 2004.09514}}].

\bibitem{Kristjansen:2023ysz}
C.~Kristjansen and K.~Zarembo, \emph{{\textquoteright{}t Hooft loops and
  integrability}}, \href{https://doi.org/10.1007/JHEP08(2023)184}{\emph{JHEP}
  {\bfseries 08} (2023) 184}
  [\href{https://arxiv.org/abs/2305.03649}{{\ttfamily 2305.03649}}].

\bibitem{Kristjansen:2021abc}
C.~Kristjansen, D.-L. Vu and K.~Zarembo, \emph{{Integrable domain walls in ABJM
  theory}}, \href{https://doi.org/10.1007/JHEP02(2022)070}{\emph{JHEP}
  {\bfseries 02} (2022) 070}
  [\href{https://arxiv.org/abs/2112.10438}{{\ttfamily 2112.10438}}].

\bibitem{Gombor:2022aqj}
T.~Gombor and C.~Kristjansen, \emph{{Overlaps for matrix product states of
  arbitrary bond dimension in ABJM theory}},
  \href{https://doi.org/10.1016/j.physletb.2022.137428}{\emph{Phys. Lett. B}
  {\bfseries 834} (2022) 137428}
  [\href{https://arxiv.org/abs/2207.06866}{{\ttfamily 2207.06866}}].

\bibitem{Jiang:2019zig}
Y.~Jiang, S.~Komatsu and E.~Vescovi, \emph{{Exact Three-Point Functions of
  Determinant Operators in Planar $N=4$ Supersymmetric Yang-Mills Theory}},
  \href{https://doi.org/10.1103/PhysRevLett.123.191601}{\emph{Phys. Rev. Lett.}
  {\bfseries 123} (2019) 191601}
  [\href{https://arxiv.org/abs/1907.11242}{{\ttfamily 1907.11242}}].

\bibitem{Jiang:2019xdz}
Y.~Jiang, S.~Komatsu and E.~Vescovi, \emph{{Structure constants in $
  \mathcal{N} $ = 4 SYM at finite coupling as worldsheet g-function}},
  \href{https://doi.org/10.1007/JHEP07(2020)037}{\emph{JHEP} {\bfseries 07}
  (2020) 037} [\href{https://arxiv.org/abs/1906.07733}{{\ttfamily
  1906.07733}}].

\bibitem{Caetano:2020dyp}
J.~a. Caetano and S.~Komatsu, \emph{{Functional equations and separation of
  variables for exact $g$-function}},
  \href{https://doi.org/10.1007/JHEP09(2020)180}{\emph{JHEP} {\bfseries 09}
  (2020) 180} [\href{https://arxiv.org/abs/2004.05071}{{\ttfamily
  2004.05071}}].

\bibitem{Caetano:2021dbh}
J.~Caetano and S.~Komatsu, \emph{{Crosscap States in Integrable Field Theories
  and Spin Chains}}, \href{https://doi.org/10.1007/s10955-022-02914-6}{\emph{J.
  Statist. Phys.} {\bfseries 187} (2022) 30}
  [\href{https://arxiv.org/abs/2111.09901}{{\ttfamily 2111.09901}}].

\bibitem{Yang:2021hrl}
P.~Yang, Y.~Jiang, S.~Komatsu and J.-B. Wu, \emph{{Three-point functions in
  ABJM and Bethe Ansatz}},
  \href{https://doi.org/10.1007/JHEP01(2022)002}{\emph{JHEP} {\bfseries 01}
  (2022) 002} [\href{https://arxiv.org/abs/2103.15840}{{\ttfamily
  2103.15840}}].

\bibitem{Fioretto:2009yq}
D.~Fioretto and G.~Mussardo, \emph{{Quantum Quenches in Integrable Field
  Theories}}, \href{https://doi.org/10.1088/1367-2630/12/5/055015}{\emph{New J.
  Phys.} {\bfseries 12} (2010) 055015}
  [\href{https://arxiv.org/abs/0911.3345}{{\ttfamily 0911.3345}}].

\bibitem{pozsgay2014correlations}
B.~Pozsgay, M.~Mesty{\'a}n, M.~A. Werner, M.~Kormos, G.~Zar{\'a}nd and
  G.~Tak{\'a}cs, \emph{Correlations after quantum quenches in the x x z spin
  chain: Failure of the generalized gibbs ensemble}, {\emph{Physical review
  letters} {\bfseries 113} (2014) 117203}.

\bibitem{Piroli:2017sei}
L.~Piroli, B.~Pozsgay and E.~Vernier, \emph{{What is an integrable quench?}},
  \href{https://doi.org/10.1016/j.nuclphysb.2017.10.012}{\emph{Nucl. Phys. B}
  {\bfseries 925} (2017) 362}
  [\href{https://arxiv.org/abs/1709.04796}{{\ttfamily 1709.04796}}].

\bibitem{Piroli:2018ksf}
L.~Piroli, E.~Vernier, P.~Calabrese and B.~Pozsgay, \emph{{Integrable quenches
  in nested spin chains I: the exact steady states}},
  \href{https://doi.org/10.1088/1742-5468/ab1c51}{\emph{J. Stat. Mech.}
  {\bfseries 1906} (2019) 063103}
  [\href{https://arxiv.org/abs/1811.00432}{{\ttfamily 1811.00432}}].

\bibitem{piroli2019integrable}
L.~Piroli, E.~Vernier, P.~Calabrese and B.~Pozsgay, \emph{Integrable quenches
  in nested spin chains ii: fusion of boundary transfer matrices},
  {\emph{Journal of Statistical Mechanics: Theory and Experiment} {\bfseries
  2019} (2019) 063104}.

\bibitem{Gombor:2021uxz}
T.~Gombor and B.~Pozsgay, \emph{{On factorized overlaps: Algebraic Bethe
  Ansatz, twists, and Separation of Variables}},
  \href{https://doi.org/10.1016/j.nuclphysb.2021.115390}{\emph{Nucl. Phys. B}
  {\bfseries 967} (2021) 115390}
  [\href{https://arxiv.org/abs/2101.10354}{{\ttfamily 2101.10354}}].

\bibitem{pozsgay2019integrable}
B.~Pozsgay, L.~Piroli and E.~Vernier, \emph{Integrable matrix product states
  from boundary integrability}, {\emph{SciPost Physics} {\bfseries 6} (2019)
  062}.

\bibitem{deLeeuw:2016umh}
M.~de~Leeuw, C.~Kristjansen and S.~Mori, \emph{{AdS/dCFT one-point functions of
  the SU(3) sector}},
  \href{https://doi.org/10.1016/j.physletb.2016.10.044}{\emph{Phys. Lett. B}
  {\bfseries 763} (2016) 197}
  [\href{https://arxiv.org/abs/1607.03123}{{\ttfamily 1607.03123}}].

\bibitem{DeLeeuw:2019ohp}
M.~De~Leeuw, T.~Gombor, C.~Kristjansen, G.~Linardopoulos and B.~Pozsgay,
  \emph{{Spin Chain Overlaps and the Twisted Yangian}},
  \href{https://doi.org/10.1007/JHEP01(2020)176}{\emph{JHEP} {\bfseries 01}
  (2020) 176} [\href{https://arxiv.org/abs/1912.09338}{{\ttfamily
  1912.09338}}].

\bibitem{Gombor:2021hmj}
T.~Gombor, \emph{{On exact overlaps for gl(N) symmetric spin chains}},
  \href{https://doi.org/10.1016/j.nuclphysb.2022.115909}{\emph{Nucl. Phys. B}
  {\bfseries 983} (2022) 115909}
  [\href{https://arxiv.org/abs/2110.07960}{{\ttfamily 2110.07960}}].

\bibitem{Gombor:2022deb}
T.~Gombor, \emph{{Integrable crosscap states in $ \mathfrak{gl} $(N) spin
  chains}}, \href{https://doi.org/10.1007/JHEP10(2022)096}{\emph{JHEP}
  {\bfseries 10} (2022) 096}
  [\href{https://arxiv.org/abs/2207.10598}{{\ttfamily 2207.10598}}].

\bibitem{Gombor:2023bez}
T.~Gombor, \emph{{Exact overlaps for all integrable two-site boundary states of
  $\mathfrak{gl}(N)$ symmetric spin chains}},
  \href{https://arxiv.org/abs/2311.04870}{{\ttfamily 2311.04870}}.

\bibitem{Sklyanin:1984sb}
E.~K. Sklyanin, \emph{{The Quantum Toda Chain}}, {\emph{Lect. Notes Phys.}
  {\bfseries 226} (1985) 196}.

\bibitem{Maillet:2018bim}
J.~M. Maillet and G.~Niccoli, \emph{{On quantum separation of variables}},
  \href{https://doi.org/10.1063/1.5050989}{\emph{J. Math. Phys.} {\bfseries 59}
  (2018) 091417} [\href{https://arxiv.org/abs/1807.11572}{{\ttfamily
  1807.11572}}].

\bibitem{Ryan:2018fyo}
P.~Ryan and D.~Volin, \emph{{Separated variables and wave functions for
  rational gl(N) spin chains in the companion twist frame}},
  \href{https://doi.org/10.1063/1.5085387}{\emph{J. Math. Phys.} {\bfseries 60}
  (2019) 032701} [\href{https://arxiv.org/abs/1810.10996}{{\ttfamily
  1810.10996}}].

\bibitem{Ryan:2020rfk}
P.~Ryan and D.~Volin, \emph{{Separation of Variables for Rational $\mathfrak
  {gl}(\mathsf {n})$ Spin Chains in Any Compact Representation, via Fusion,
  Embedding Morphism and B\"acklund Flow}},
  \href{https://doi.org/10.1007/s00220-021-03990-7}{\emph{Commun. Math. Phys.}
  {\bfseries 383} (2021) 311}
  [\href{https://arxiv.org/abs/2002.12341}{{\ttfamily 2002.12341}}].

\bibitem{Frassek:2011aa}
R.~Frassek, T.~Lukowski, C.~Meneghelli and M.~Staudacher, \emph{{Baxter
  Operators and Hamiltonians for 'nearly all' Integrable Closed
  $\mathfrak{gl}(n)$ Spin Chains}},
  \href{https://doi.org/10.1016/j.nuclphysb.2013.06.006}{\emph{Nucl. Phys. B}
  {\bfseries 874} (2013) 620}
  [\href{https://arxiv.org/abs/1112.3600}{{\ttfamily 1112.3600}}].

\bibitem{Kazakov:2015efa}
V.~Kazakov, S.~Leurent and D.~Volin, \emph{{T-system on T-hook: Grassmannian
  Solution and Twisted Quantum Spectral Curve}},
  \href{https://doi.org/10.1007/JHEP12(2016)044}{\emph{JHEP} {\bfseries 12}
  (2016) 044} [\href{https://arxiv.org/abs/1510.02100}{{\ttfamily
  1510.02100}}].

\bibitem{Gromov:2016itr}
N.~Gromov, F.~Levkovich-Maslyuk and G.~Sizov, \emph{{New Construction of
  Eigenstates and Separation of Variables for SU(N) Quantum Spin Chains}},
  \href{https://doi.org/10.1007/JHEP09(2017)111}{\emph{JHEP} {\bfseries 09}
  (2017) 111} [\href{https://arxiv.org/abs/1610.08032}{{\ttfamily
  1610.08032}}].

\bibitem{Maillet:2020ykb}
J.~M. Maillet, G.~Niccoli and L.~Vignoli, \emph{{On Scalar Products in Higher
  Rank Quantum Separation of Variables}},
  \href{https://doi.org/10.21468/SciPostPhys.9.6.086}{\emph{SciPost Phys.}
  {\bfseries 9} (2020) 086} [\href{https://arxiv.org/abs/2003.04281}{{\ttfamily
  2003.04281}}].

\bibitem{pozsgay2014overlaps}
B.~Pozsgay, \emph{Overlaps between eigenstates of the xxz spin-1/2 chain and a
  class of simple product states}, {\emph{Journal of Statistical Mechanics:
  Theory and Experiment} {\bfseries 2014} (2014) P06011}.

\bibitem{Ghoshal:1993tm}
S.~Ghoshal and A.~B. Zamolodchikov, \emph{{Boundary S matrix and boundary state
  in two-dimensional integrable quantum field theory}},
  \href{https://doi.org/10.1142/S0217751X94001552}{\emph{Int. J. Mod. Phys. A}
  {\bfseries 9} (1994) 3841}
  [\href{https://arxiv.org/abs/hep-th/9306002}{{\ttfamily hep-th/9306002}}].

\bibitem{molev2007yangians}
A.~Molev, \emph{Yangians and classical Lie algebras}, no.~143. American
  Mathematical Soc., 2007.

\bibitem{Maillet:2018czd}
J.~M. Maillet and G.~Niccoli, \emph{{Complete spectrum of quantum integrable
  lattice models associated to Y(gl(n)) by separation of variables}},
  \href{https://doi.org/10.21468/SciPostPhys.6.6.071}{\emph{SciPost Phys.}
  {\bfseries 6} (2019) 071} [\href{https://arxiv.org/abs/1810.11885}{{\ttfamily
  1810.11885}}].

\bibitem{Jiang:2020sdw}
Y.~Jiang and B.~Pozsgay, \emph{{On exact overlaps in integrable spin chains}},
  \href{https://doi.org/10.1007/JHEP06(2020)022}{\emph{JHEP} {\bfseries 06}
  (2020) 022} [\href{https://arxiv.org/abs/2002.12065}{{\ttfamily
  2002.12065}}].

\bibitem{Frassek:2017fba}
R.~Frassek, \emph{{Boundary Perimeter Bethe Ansatz}},
  \href{https://doi.org/10.1088/1751-8121/aa7278}{\emph{J. Phys. A} {\bfseries
  50} (2017) 265202} [\href{https://arxiv.org/abs/1703.10842}{{\ttfamily
  1703.10842}}].

\bibitem{Ekman:2022kpx}
C.~Ekman, \emph{{Crosscap states in the XXX spin-1/2 spin chain}},
  \href{https://arxiv.org/abs/2207.12354}{{\ttfamily 2207.12354}}.

\bibitem{Widen:2018nnu}
E.~Wid\'en, \emph{{One-point functions in $\beta$-deformed $ \mathcal{N}=4 $
  SYM with defect}}, \href{https://doi.org/10.1007/JHEP11(2018)114}{\emph{JHEP}
  {\bfseries 11} (2018) 114}
  [\href{https://arxiv.org/abs/1804.09514}{{\ttfamily 1804.09514}}].

\bibitem{pozsgay2018overlaps}
B.~Pozsgay, \emph{Overlaps with arbitrary two-site states in the xxz spin
  chain}, {\emph{Journal of Statistical Mechanics: Theory and Experiment}
  {\bfseries 2018} (2018) 053103}.

\bibitem{Kristjansen:2020vbe}
C.~Kristjansen, D.~M\"uller and K.~Zarembo, \emph{{Overlaps and fermionic
  dualities for integrable super spin chains}},
  \href{https://doi.org/10.1007/JHEP03(2021)100}{\emph{JHEP} {\bfseries 03}
  (2021) 100} [\href{https://arxiv.org/abs/2011.12192}{{\ttfamily
  2011.12192}}].

\bibitem{Kristjansen:2021xno}
C.~Kristjansen, D.~M\"uller and K.~Zarembo, \emph{{Duality relations for
  overlaps of integrable boundary states in AdS/dCFT}},
  \href{https://doi.org/10.1007/JHEP09(2021)004}{\emph{JHEP} {\bfseries 09}
  (2021) 004} [\href{https://arxiv.org/abs/2106.08116}{{\ttfamily
  2106.08116}}].

\bibitem{Buhl-Mortensen:2017ind}
I.~Buhl-Mortensen, M.~de~Leeuw, A.~C. Ipsen, C.~Kristjansen and M.~Wilhelm,
  \emph{{Asymptotic One-Point Functions in Gauge-String Duality with Defects}},
  \href{https://doi.org/10.1103/PhysRevLett.119.261604}{\emph{Phys. Rev. Lett.}
  {\bfseries 119} (2017) 261604}
  [\href{https://arxiv.org/abs/1704.07386}{{\ttfamily 1704.07386}}].

\bibitem{Shahpo:2021xax}
O.~Shahpo and E.~Vescovi, \emph{{Correlation functions of determinant operators
  in conformal fishnet theory}},
  \href{https://doi.org/10.1007/JHEP06(2022)070}{\emph{JHEP} {\bfseries 06}
  (2022) 070} [\href{https://arxiv.org/abs/2110.09458}{{\ttfamily
  2110.09458}}].

\bibitem{Ryan:2021duf}
P.~Ryan, \emph{{Integrable systems, separation of variables and the Yang-Baxter
  equation}}, Ph.D. thesis, TCD, Dublin, 2021.
\newblock \href{https://arxiv.org/abs/2201.12057}{{\ttfamily 2201.12057}}.

\end{thebibliography}\endgroup

\end{document}